\documentclass[useAMS,usenatbib]{mnras}
\usepackage{graphicx,float,subfigure} 
\usepackage[export]{adjustbox}
\usepackage{array,longtable}
\usepackage{natbib}
\usepackage{float}
\usepackage[T1]{fontenc}
\usepackage{ae}
\usepackage{aecompl}
\usepackage{amsmath, amssymb, amsfonts, braket, gensymb} %math symbols/writing

\usepackage{multicol}
\usepackage{newtxtext,newtxmath} %Times font
\usepackage{lipsum}

\def\url#1{\expandafter\string\csname #1\endcsname}

\def\actaa{Acta. Astronom.} 				%
\def\aj{AJ} 								% Astronomical Journal
\def\araa{ARA\&A} 							% Annual Review of Astron and Astrophys
\def\apj{ApJ}	 							% Astrophysical Journal
\def\apjl{ApJ}	 							% Astrophysical Journal, Letters
\def\apjs{ApJS}        						% Astrophysical Journal, Supplement
\def\ao{Appl.~Opt.}    						% Applied Optics
\def\aap{A\&A}         						% Astronomy and Astrophysics
\def\aaps{A\&AS}       						% Astronomy and Astrophysics, Supplement
\def\mnras{MNRAS}       					% Monthly Notices of the RAS
\def\pasa{PASA}								% Publications of the Astronomical Society of Australia
\def\pasp{PASP}		       					% Publications of the ASP
\def\nat{Nature}	        				% Nature
\def\memsai{Mem.~Soc.~Astron.~Italiana}     % Mem. Societa Astronomica Italiana
\title[Helium-enriched globular cluster progenitors]{The Predicted Properties of Helium-Enriched Globular Cluster Progenitors at High Redshift}
\author[Nataf et al.]{David M. Nataf$^1$\thanks{Email: dnataf1@jhu.edu, david.nataf@gmail.com}, Shunsaku Horiuchi$^2$, Guglielmo Costa$^{3,4}$, 
\newauthor {Rosemary F. G. Wyse$^1$, Yuan-Sen Ting$^{5,6,7}$, Roland Crocker$^{8}$, }
\newauthor {Christoph Federrath$^{8}$, Yang Chen$^{4}$\thanks{Email: yang.chen@unipd.it} } 
\vspace*{6pt}\\
$^1$Department of Physics and Astronomy, The Johns Hopkins University, Baltimore, MD 21218, \\
$^{2}$Center for Neutrino Physics, Department of Physics, Virginia Tech, Blacksburg, VA
    24061, USA \\
$^{3}$SISSA - ISAS - International School for Advanced Studies, Via Bonomea 265, 34136, Trieste, Italy  \\
$^{4}$Dipartimento di Fisica e Astronomia Galileo Galilei, Universit\`a di Padova, Vicolo dell'Osservatorio 3, I-35122 Padova, Italy \\    
$^5$Institute for Advanced Study, Princeton, NJ 08540, USA \\
$^6$Department of Astrophysical Sciences, Princeton University, Princeton, NJ 08544, USA \\
$^7$Observatories of the Carnegie Institution of Washington, 813 Santa Barbara Street, Pasadena, CA 91101, USA \\
$^{8}$Research School of Astronomy and Astrophysics, Australian National University, Canberra, ACT 2611, Australia}

\begin{document}
\date{Accepted ...... Received ...... ; in original form......   }
\pagerange{\pageref{firstpage}--\pageref{lastpage}} \pubyear{2018}
\maketitle
\label{firstpage}

\begin{abstract}
Globular cluster progenitors may have been detected by \textit{HST}, and are predicted to be observable with \textit{JWST} and ground-based extremely-large telescopes with adaptive optics. This has the potential to elucidate the issue of globular cluster formation and the origins of significantly helium-enriched subpopulations, a problem in Galactic astronomy with no satisfactory theoretical solution. Given this context, we use model stellar tracks and isochrones to investigate the predicted observational properties of helium-enriched stellar populations in globular cluster progenitors. We find that, relative to helium-normal populations, helium-enriched (${\Delta}Y=+0.12$) stellar populations similar to those inferred in the most massive globular clusters, are expected, modulo some rapid fluctuations in the first $\sim$30 Myr,  to be brighter and redder in the rest frame. At fixed age, stellar mass, and metallicity, a helium-enriched population is predicted to converge to being $\sim$0.40 mag brighter at $\lambda \approx 2.0\, {\mu}m$, and to be 0.30 mag redder in the \textit{JWST}-NIRCam colour $(F070W-F200W)$, and to actually be fainter for $\lambda \lesssim 0.50 \, {\mu}m$. Separately, we find that the time-integrated shift in ionizing radiation is a negligible $\sim 5\%$, though we show that the Lyman-$\alpha$ escape fraction could end up higher for helium-enriched stars. %However, it may still end up resulting in an enhancement to the Lyman-$\alpha$ photon budget, if the surrounding interstellar media are helium-enriched as well. 
\end{abstract}
\maketitle

\begin{keywords}
globular clusters: general 
\end{keywords}

%%%%% BODY TEXT %%%%%
\section{Introduction}
\label{sec:Introduction}
%Globular clusters once provided an important bridge linking Galactic astronomy and cosmology, due to the cosmic age problem. The inferred ages for Galactic globular clusters ran as high as $\sim$18 Gyr \citep{1992ApJS...81..163B}, and thus implausibly older than the inferred age of the universe itself. Overtime, this was resolved due to a combination of factors, such as an improved distance scale to globular clusters \citep{1998ApJ...494...96C}, and improved stellar models which incorporated physics such as updated opacity tables \citep{1997ApJ...479..665S}. At the other end, the inferred age of the universe was increased due to reduced measured values of Hubble's constant \citep{2001ApJ...553...47F}, and the detection of a cosmological constant \citep{1998AJ....116.1009R,1999ApJ...517..565P}. Though this problem is resolved, a new bridge involving globular clusters may be emerging, this time linking Galactic and high-redshift astronomy through to the reionization epoch. 

Globular cluster populations are predicted to trace the formation histories of galaxies. That is because mergers can both spur the formation of new globular clusters, and bring in additional clusters as part of hierarchical assembly \citep{2019MNRAS.486.3134K,2019MNRAS.486.3180K}.

\citet{1992ApJ...384...50A} and \citet{1993MNRAS.264..611Z}  proposed that the high pressure in the interstellar medium of merging, gas-rich systems could explain the mass-radius relation of old globular clusters, and the number and luminosity of young star clusters in currently merging systems. %They also showed that merger histories  could explain the presence of two peaks in the distribution of the mean metallicities of globular clusters. 
A supporting example is that of the major merging system Arp 220, which hosts at least 24 young star clusters with masses spanning a similar range as that of the Milky Way's 150+ globular clusters \citep{2006ApJ...641..763W}. More recently, the semi-analytic model of \citet{2010ApJ...718.1266M} predicted that early mergers of smaller hosts create exclusively blue clusters, whereas subsequent mergers of more massive galaxies create both red and blue clusters, with the fraction of galactic stellar mass being in globular clusters declining from $\sim$10\% at high redshift to $\sim 0.1\%$ at present. \citet{2018ApJ...861..148M} used cosmological zoom-in simulations to propose that $\sim 30 \%$ of the Milky Way's metal-poor globular clusters originated from cold filamentary accretion, with the remainder accreted from mergers. Their scenario, which they predict to be most pronounced at a time corresponding to $z \sim 6$\footnote{Throughout this work, lower case "z" is used to denote cosmological redshift, and uppercase "Z" to denote the initial metals fraction of stars.},  can account for the observed kinematics and spatial distribution of metal-poor globular clusters. 

Globular clusters may also have significantly contributed to reionization.  \citet{2017MNRAS.472.3120B,2018MNRAS.479..332B} estimated that, for reasonable assumptions of the escape fraction, no less than 20\% of the ionizing sources with $M_{1500 \AA} \approx -12$ may have been globular clusters at $z \approx 4$. \citet{2018MNRAS.477..480Z} have shown that for the progenitors of systems such as the Fornax dwarf galaxy, the globular clusters could account for a far greater share of their UV flux than their stellar mass fraction, if observed at the right times, since individual massive globular cluster stellar populations are formed in short-time intervals. \citet{2018MNRAS.477..480Z} also showed that ignoring this and related features can induce order-of-magnitude errors in abundance matching, between the first halos to form stars and the galaxies observed in high-redshift data. 
Thus, a better understanding of globular cluster formation may be necessary to complete an inventory of ionizing photons during the reionization epoch. 

Though the value of 20\% of the ionizing sources being globular clusters satisfies the trivial constraint of being less than 100\%, it is still disconcertingly high, as one can argue for changes to two of the assumptions made by \citet{2018MNRAS.479..332B}. These are that globular cluster stellar masses were not greater at the time of their birth than their present-day masses, and that there were no globular clusters that have since completely disassociated. These assumptions are not physically valid \citep{1987degc.book.....S,2001ApJ...561..751F}, and thus the 20\% estimate is lower than a proper lower bound. That can be seen in Figure 5 of  \citet{2018MNRAS.477..480Z}, and is acknowledged in several sections of that work. If the initial masses of globular cluster were merely $\sim 10 \times$ higher, then their ionizing flux would exceed the measured luminosity function over the magnitude range $-19 \lesssim M_{1500} \lesssim -12$, where $M_{1500}$ is the absolute magnitude of a source at $\lambda = 1500$~\AA. 

\citet{2009ApJ...698L.158K} estimated on dynamical grounds that the total stellar mass once formed in globular clusters is $\sim 40 \times$ greater than their current total mass, with the losses due to both some globular clusters having lost some mass, and many having been completely disassociated. Renormalizing the 20\% estimate by this mass loss argument shows that the estimates of \citet{2009ApJ...698L.158K} and \citet{2018MNRAS.479..332B} are mutually exclusive, at a factor $\sim$10 level, if we assume the ionizing radiation field estimates of \citep{2016PASA...33...37F}. Separately, \citet{2017MNRAS.471.3668S} have measured the present-day stellar mass functions of 35 Galactic globular clusters spanning the range $dn/dm \propto m^{-1.89}$ to $m^{+0.11}$, with a median of $m^{-0.69}$. Their measured mass functions are tightly correlated with the ratio of the age to the half-mass relaxation times of the clusters, where that ratio can be called a \textit{dynamical age}. That is the expectation if shallower mass functions originate from dynamical evaporation, in which the lower-mass stars are expected to be preferentially lost over time due to mass segregation, super-imposed on a universal initial mass function \citep{2003MNRAS.340..227B,2013MNRAS.433.1378L,2015MNRAS.453.3278W}. Using these constraints, \citet{2017MNRAS.471.3668S} estimate that $\sim 2 \times 10^{8} M_{\odot}$ of Milky Way stars are evaporated members of the surviving globular clusters, which is some $\sim 7 \times$ greater than the current integrated stellar mass of the Milky Way globular cluster system \citep{2009ApJ...698L.158K}, and comparable to the current literature consensus estimate for the total stellar mass of the halo of $(4 - 7) \times 10^{8} M_{\odot}$ \citep{2016ARA&A..54..529B}. 

These dynamical arguments are now somewhat supported by chemical evolution arguments. Globular clusters host multiple populations differing in chemistry, which are typically assumed to be "generations".  Typically, some $\sim 1/3$ of the stars in globular clusters have abundances that are consistent with the metallicity-dependent trends of the Milky Way Halo, Thick Disk, and Bulge \citep{2009A&A...505..117C}, whereas the remaining two thirds show various light-element trends, such as enrichment of sodium and almuinium and depletion of oxygen and magnesium \citep{2009A&A...505..117C,2017AJ....154..155J,2017ApJ...836..168J,2019AJ....158...14N}, and enrichment in helium \citep{2004ApJ...612L..25N,2007ApJ...661L..53P}. It is conventionally assumed -- but not demonstrated -- that the population showing anomalous abundances formed some unspecified time after the population showing normal abundances, from the gaseous ejecta of the first generation stars. These form a ``first" and a ``second" generation. The chemical properties of multiple populations are correlated with the stellar mass and metallicity of the host globular clusters \citep{2017MNRAS.464.3636M,2019AJ....158...14N}. 

If one assumes a standard initial stellar mass function and chemical yield values, the surviving globular clusters must have been no less than $\sim 10 \times$ more massive at birth \citep{2011A&A...533A.120V,2012ApJ...758...21C}. That estimate results from requiring the ejecta of the asymptotic giant branch stars of the first generations to have sufficient mass to be recycled as the second generation, with its present-day number counts. This is referred to as the ``mass-budget problem". There is currently no model in the literature  that succeeds at accounting for each of the dynamical constraints, the abundance trends, and the mass-budget problem \citep{2013MmSAI..84..162R,2018ARA&A..56...83B}. Though both dynamical and chemical arguments necessitate globular cluster mass loss, the quantity of mass loss has not been shown to be commensurate. There is no viable, general theory of globular cluster formation at this time, just a sea of unexplained observations and desperate phenomenology. 

\citet{2010A&A...519A..14M} and \citet{2017MNRAS.465..501S} used chemical abundance measurements of field stars to argue that a significant fraction, perhaps $\sim$50\%, of  Milky Way stars with [Fe/H] $\leq -1$ originated in globular clusters hosting multiple generations. That is in contrast to star formation today, which takes place in lower-mass associations and open clusters. These do not show the abundance anomalies associated with globular clusters \citep{2014ApJ...796...68B,2015ApJ...798L..41C,2018A&A...619A.176B}. It has not been ascertained if this is predominantly due to stars ejected from surviving globular clusters, or from now fully disassociated globular clusters. That should eventually be straightforward to measure as the precision of chemodynamical tagging improves \citep{2015ApJ...807..104T}. 

%At the typical age of globular clusters, $\sim$12 Gyr \citep{2009ApJ...694.1498M}

Perhaps surprisingly, searches for the relevant abundance anomalies \citep{2016MNRAS.460.1869C,2017MNRAS.468.3150M} or age spreads \citep{2014MNRAS.441.2754C} in young star clusters in nearby locations such as the Large Magellanic Cloud with mass and size comparable to those of present-day Galactic globular clusters have yielded null results. The creation of multiple stellar generations in globular clusters may be a feature of the high-redshift universe. It has been demonstrated that young clusters do host stars with a large and unexplained spread in angular momentum \citep{2016MNRAS.460L..20B,2018MNRAS.480.3739B,2018AJ....156..116M,2019ApJ...876...65L}, but at this time it is unknown if that phenomenon is related to that of chemically distinct populations in massive globular clusters. Separately, the theoretical investigation of \citet{2020MNRAS.494.3861R} showed that more rapidly rotating massive stars are expected to experience more and earlier dredge up, and thus shift to redder colours earlier in their evolution. It is also the case that lower-metallicity massive stars are expected to have much faster rotation rates \citep{2013AN....334..595C}. Given these factors, a separate study investigating the predicted effects of rotation on young massive clusters would be worthwhile. 

In that case, directly constraining the nature of these systems' formation can only come from deep, precise imaging and spectroscopy.  \citet{2017MNRAS.469L..63R} has shown that globular cluster progenitors should be detectable with the \textit{James Webb Space Telescope (JWST)} in each of the optical, near-IR, and mid-IR bands, even with modest exposure times of 10,000 seconds, or approximately 3 hours. Deep observations can establish independent constraints on the initial masses of globular clusters, their formation history as a function of redshift, and the physical state of their host galaxies at the time of globular cluster formation, with the total numbers of globular clusters detected being a sensitive indicator of the initial mass function of globular clusters \citep{2019MNRAS.485.5861P}.  \citet{2019MNRAS.485.5861P} computed more detailed estimates, and concluded that the total number counts of observed globular cluster progenitors would sharply constrain their initial mass function, independently of the formation redshift. We note that the first suggestion of such observations was that of \citet{2002ApJ...573...60C}, who predicted the luminosity function and clustering of globular cluster progenitors at high redshift. 

The proposal that \textit{JWST} (and eventually extremely large ground-based telescopes with adaptive optics) might constrain globular cluster formation models is supported by arguments that globular cluster progenitors might already have been observed with the \textit{Hubble Space Telescope (HST)} \citep{2017arXiv171102090B,2019MNRAS.483.3618V}. \citet{2019arXiv190407941V} have actually obtained spectra for one of these candidate clusters. It is at a redshift $z=3.121$, has an estimated stellar mass $M \lesssim 10^{7} M_{\odot}$, an effective radius smaller than 20 pc, and does indeed punch a hole through the surrounding medium to release some ionizing radiation. 

%It is therefore reasonable to expect scientific cross-pollination between high-redshift observations and Galactic archaeology driven by globular clusters in the era of \textit{JWST} Inferences on the history of Galactic globular clusters are interesting: they host multiple stellar generations \citep{2004ApJ...612L..25N,2007ApJ...661L..53P}, which are typically distinct in light elements (He, Na, O, etc) but not usually iron \citep{2009A&A...505..117C}, they're ancient with typical ages of $\sim$ 12 Gyr \citep{2009ApJ...694.1498M}, and their progenitors may be responsible for the bulk of Milky Way star formation for stars with [Fe/H] $\leq -1$ \citep{2010A&A...519A..14M,2017MNRAS.465..501S}. These constraints, though well-founded, cannot be matched by any current formation model, due to both the issue of missing mass and the peculiarities of the chemical abundance anomalies \citep{2013MmSAI..84..162R,2018ARA&A..56...83B}.  Nevertheless, these constraints are robust, and derived from numerous, independent, state-of-the-art observational programs. 

In this paper we investigate the predicted observable quantities of helium enrichment on future, high-redshift observations of globular cluster progenitors. We assume that globular clusters do not just undergo a single starburst, but typically two or more. The second generation is often enriched in helium, with the most massive globular clusters typically hosting second stellar generations with the largest enrichments in helium ( e.g. \citealt{2004ApJ...612L..25N,2007ApJ...661L..53P,2017MNRAS.464.3636M}). The most massive globular clusters are also expected to be the most easily observable \citep{2017MNRAS.469L..63R,2019MNRAS.485.5861P}, and it thus follows the most easily observable clusters will likely be those with higher stellar helium abundances. We thus explore and investigate the effect of helium enrichment on the evolution of high-mass stars using the best-available tracks and isochrones for this purpose, in combination with modelling of synthetic simple stellar populations. 

The structure of this paper is as follows. %We further motivate our investigation of the effects of helium-enrichment in globular clusters in Section \ref{sec:MassHelium}. 
In Section \ref{sec:StellarModels} we review some of the applications of stellar models in interpreting the properties of helium-enriched populations. In Section \ref{sec:HeliumPrediction} and  \ref{sec:HeliumPredictionStarbursts}, we explore the effect of helium enrichment on stellar tracks of high-mass stars, and on combined simple stellar populations, respectively, which are due to helium-dependent shifts in the lifetimes of stars and their evolutionary tracks in the  Hertzprung-Russel diagram. In a brief segue in Section \ref{sec:OtherEffects}, we explore the possibility that other factors may vary, such as the escape fraction of ioninizing photons. We conclude in Section \ref{sec:Conclusion}.

\section{Helium-enriched Stellar Models}
\label{sec:StellarModels}

\subsection{Theoretical primer}
\label{subsec:TheoreticalPrimer}

In the subsequent subsections we will discuss various specific and detailed predictions of helium-enriched stellar models. Before doing so, however, it is worth noting that the \textit{qualitative} predictions have long been understood. It was \citet{1952ApJ...115..326S} who discovered that helium fusion could be an important energy source for stars that had depleted their hydrogen, and \citet{1953PASP...65..210C} who subsequently published the first stellar models for pure helium stars. 

Helium-enriched stars are expected to be different from helium-normal stars due to, among other reasons:
\begin{itemize}
    \item Being born with their main fuel source (hydrogen) being already partially depleted;
    \item Their greater mean molecular weight, which makes nuclear fusion more efficient;
    \item Their reduced opacity, as helium has fewer electrons than hydrogen at fixed total mass;
\end{itemize}
These issues are discussed and explained in many standard textbooks on stellar structure and evolution (e.g. \citealt{1990sse..book.....K,1992itsa.book.....B,2005essp.book.....S}).

\subsection{Previous theoretical work on helium-enriched stellar populations}
\label{subsec:PreviousModels}

The effects of helium-enrichment on stellar models, largely driven by the need to model and understand observations of present-day Galactic globular clusters, has been predominantly discussed and explored at lower stellar masses. Among these:
\begin{itemize}
\item At fixed age and metallicity, helium-enriched main-sequence and red giant branch populations are predicted to be bluer, with a dimmer main-sequence turnoff \citep{2008ApJS..178...89D}. 
%%%%%%%%%%
%\item Given that prediction that helium-enriched populations should have bluer main-sequences at fixed age and metallicity, the colour trend of solar neighbourhood K-dwarfs with metallicity should constrain the helium-to-metals enrichment ratio ${\Delta}Y/{\Delta}Z$. K-dwarfs are selected for this method as they have relatively weak trends with age. However, the method fails and yields an unphysical result of ${\Delta}Y/{\Delta}Z \approx 5.3$, extrapolating to values $Y < Y_{\rm{BBN}}$ for low metallicities \citep{2007MNRAS.382.1516C,2010A&A...518A..13G}. The origin of this failure is not understood at this time. 
%%%%%%%%%%%
 \item  Underestimating the initial helium abundance is predicted to lead to an overestimate of the photometrically derived age. That error is due to the combination of two factors: A helium-enriched stellar population will have a brighter horizontal branch, and a fainter main-sequence turnoff, and thus a larger difference in brightness between the two \citep{2010ApJ...714.1072M}.
 %%%%%%%%%%%%%
 \item  Increased initial helium abundance is predicted to decrease the mass of red giant branch stars at fixed age and metallicity. The functional dependence between these parameters was parameterized by \citet{2012AcA....62...33N} as $\log{\{M_{\rm{RGB}}/M_{\odot}\}} = 0.026 + 0.126\rm{[M/H]} -0.276\log{\{t/(10\,\,\rm{Gyr})\}}-0.937(Y-0.27)$.
 %%%%%%%%%%%%%%% 
\item Helium-enriched stellar populations are predicted to have a brighter \citep{1997MNRAS.285..593C,2011ApJ...736...94N,2018MNRAS.475.4088L} and less populous \citep{2001ApJ...546L.109B,2014MNRAS.445.3839N} red giant branch bump. The predicted difference in brightness is now robustly confirmed \citep{2018MNRAS.475.4088L,2019ApJ...871..140L}, but not that of number counts. That is plausibly because the latter would require a larger sample than is available.
%%%%%%%%%%%%%%%%%
\item Helium-enriched populations tend to have a bluer horizontal branch morphology at fixed age and metallicity. That is almost certainly part of the solution \citep{1994ApJ...423..248L,2006MmSAI..77..144P} to the "second parameter problem" \citep{1967ApJ...150..469S,1967AJ.....72...70V}, whereby globular clusters of similar metallicity can show very different horizontal branch morphologies. 
%%%%%%%%%%%%%%%%%%
\item Asymptotic giant branch stars that are helium-enriched and that have initial masses of $\sim1-6\,M_{\odot}$ are predicted to undergo a different number of thermal pulses, produce different chemical yields, produce fewer carbon stars, and leave behind larger-mass remnants at fixed metallicity and initial mass \citep{2014ApJ...784...32K,2014MNRAS.445..347K,2015MNRAS.452.2804S}. 
\item At fixed age, the turnoff mass of a stellar population will be lower, and thus the brightest white dwarfs in a helium-enriched population will be less massive than those of a stellar population with the same age and metallicity \citep{2017A&A...602A..13C}.
\item The larger number of white dwarfs, due to the turnoff mass being lower at fixed age, is predicted to contribute to an increase in the near-infrared mass-to-light ratio of old populations relative to scaled solar analogues. \citet{2018MNRAS.478.2368C} estimated the offset to be as high as 16\% at $\lambda \approx $ 8,500 \AA, for an extreme population with $Y=0.40$.
%\item  The shifts in the mass and number of white dwarf remnants contributed to the prediction of, that an old, helium-enriched stellar population may have a mass-to-light ratio in near-infrared that will be 20 per cent higher than that of the analogous scaled-solar population.
%%%%%%%%%%%%%%%%
\item Very massive helium-enriched stars are also expected to leave behind more massive remnants at fixed initial mass. \citet{2017MmSAI..88..244K} predict that helium-enriched ($Y=0.40$) stars with initial masses of 6~$M_{\odot}$ and 18~$M_{\odot}$ are respectively sufficient to produce neutron star and black hole remnants. Without helium-enrichment, the respective values are 8~$M_{\odot}$ and 23~$M_{\odot}$. 
\end{itemize} 

\subsection{Helium-enriched stellar models used in this investigation}
\label{subsec:TheseModels}

In this investigation, we discuss the predictions for stars that are both higher-mass and helium-enriched, given the context that high-redshift globular cluster progenitors may soon be observed in integrated light by $JWST$. To the best of our knowledge, this has not been explored in the literature. We used two sets of stellar models, which were treated differently. This combination allows for a few consistency checks to be implemented in this new predictive regime. %It also allows us to increase the breadth of predictions, as neither set of models includes the full set of relevant outputs. 

We primarily use a second set of models based on the PARSEC v1.2 code \citep{2012MNRAS.427..127B,2014MNRAS.445.4287T,2015MNRAS.452.1068C}. A grid of stellar tracks densely samples the parameter of initial stellar masses up to $M_{i} = 300 M_{\odot}$, with bolometric corrections computed following the prescription of \citet{2019A&A...632A.105C}. Three families of isochrones are constructed, with the metals fraction fixed to $Z=0.017$ (the solar value) but the initial helium abundance set to $Y=0.279$ (hereafter rounded to $Y=0.28$), 0.33, and 0.40. These models do not include the asymptotic giant branch phase, and thus we are using them only to further probe the predicted properties of helium-enriched stellar populations in the first $\sim$100 Myr, when the luminosities of higher mass stars dominate the flux budget. We also briefly discuss the evolution of core masses in these stars. 

The mass-loss prescription for these tracks and isochrones are described by \citet{2015MNRAS.452.1068C}. These assume the  metals mixture of \citet{2009A&A...498..877C}.  The mass-loss prescription is a function of evolutionary phase and metallicity of the stars. This includes different prescriptions for the mass loss during the blue supergiant phase \citep{2000A&A...362..295V,2001A&A...369..574V}, supergiants with $T_{\rm{eff}} \leq 12,000\,K$ \citep{1988A&AS...72..259D},  and stars in the Wolf-Rayet phase \citep{2000A&A...360..227N}. 

Integrated stellar properties are computed directly from the isochrones. We adopt a Salpeter IMF with slope $dN/dM \propto M^{\alpha}$ with $\alpha = -2.35$ \citep{1955ApJ...121..161S}. The total stellar mass is normalized to be $10^6 M_\odot$ over the mass range $0.5$--$300 M_\odot$. Following the work of \citet{2017MNRAS.469L..63R}, we assume that the stars are formed with a uniform age dispersion spanning 3 Myr. Absolute magnitudes on the Vega system are computed for the bandpasses $F170W$, $F218W$, $F255W$, $F300W$, $F336W$, $F439W$, $F450W$, $F555W$, $F606W$, $F702W$, $F814W$ filter of the \textit{WFC3 UVIS} imager on \textit{HST}, and for the $F070W$, $F090W$, $F115W$, $F150W$, $F200W$, $F277W$, $F356W$, $F444W$, $F150W2$, $F322W2$ of the \textit{NIRCam} imager on \textit{JWST}. The same 10 \textit {JWST} filters were recently selected in the related investigation of \citet{2019MNRAS.485.5861P}. The fluxes in the latter are computed both in the rest-frame, and redshifted to $z=0.10,0.20,..,0.90$ and $z=1,2,...,10$ using the cosmology calculator of \citep{2006PASP..118.1711W}. The assumed cosmology is $H_{0} = 69.6$, $\Omega_{M} = 0.286$, $\Omega_{\Lambda} = 0.714$ with no curvature \citep{2014ApJ...794..135B}.

We also make use of the stellar tracks of \citet{2008A&A...484..815B} and \citet{2009A&A...508..355B}. This second set of models is first used to enable a consistency and calibration check with the first, and also, because the broader range of input abundances available enable further tests of the effects of varying abundances. The input abundances include a broad range in the metals-mass fraction $Z$ and the helium mass fraction $Y$, of which we use $\log{Z/Z_{\odot}}=-2.23,-1.23,0,+0.37$ and $Y=0.26,0.30,0.34,0.40$. These models assume the solar metals mixture of \citet{1993oee..conf...15G}. %These sets of models were chosen as they include the post-main-sequence phases of stellar evolution, including the thermally-pulsating asymptotic giant branch phase, which provides a substantial fraction of the total luminosity of a stellar population. The predicted energy released by the thermally-pulsating asymptotic giant stars is consistent with estimates of white dwarf masses in open clusters \citep{2011ApJ...733...81B}.

To compute integrated light estimates from these populations, we use the {\tt PEGASE.2} evolutionary synthesis code \citep{1999astro.ph.12179F} to investigate integrated properties of synthetic stellar populations. We replace the default stellar evolutionary stellar tracks with the He-rich tracks of \cite{2009A&A...508..355B}\footnote{http://cdsweb.u-strasbg.fr/cgi-bin/qcat?J/A+A/508/355}, including the main sequence, horizontal branch and asymptotic giant branch tracks. %{\tt PEGASE.2} offers several options for population synthesis: we adopt an initial metallicity of $Z=0.017$, the supernova ejecta ``model B'' of \cite{1995ApJS..101..181W}, a close binary fraction of 0.05, no galactic winds, no treatment of dust extinction, and a minimum CCSN mass of $8 M_\odot$. The supernova model and minimum mass are only relevant to follow metal enrichment, which is effectively neglect by assuming an ISM metallicity of $Z=0.017$. Finally,
We adopt a Salpeter IMF with slope $dN/dM \propto M^{\alpha}$ with $\alpha = -2.35$ over the mass range $0.5$--$20 M_\odot$ to match the availability of the \cite{2009A&A...508..355B} tracks. With this setup, we follow the evolution of a stellar population of total initial mass $10^6 M_\odot$ undergoing continuous star formation rate with of $50 \, {\rm M_\odot \, yr^{-1}}$, separately for $Y=0.23$ through to $Y=0.4$. The total duration of star formation is 20,000 years,  and thus effectively instantaneous. %Predicted fluxes are computed for the bandpasses $UBVRI$ in Vega magnitudes, and for the bandpasses $vgr$ \citep{1976PASP...88..543T} in AB magnitudes.  

The main limitations to the models of \citet{2009A&A...508..355B} are the coarseness of the mass grid at high masses, for which the available masses are $M/M_{\odot}=8, 10, 12, 15, 20$, and the absence of models for stars with initial masses $M/M_{\odot} > 20$.

\section{The predicted effect of enriched helium on the evolution of single stars}
\label{sec:HeliumPrediction}

\begin{figure}
\centering
\includegraphics[width=0.45\textwidth, center]{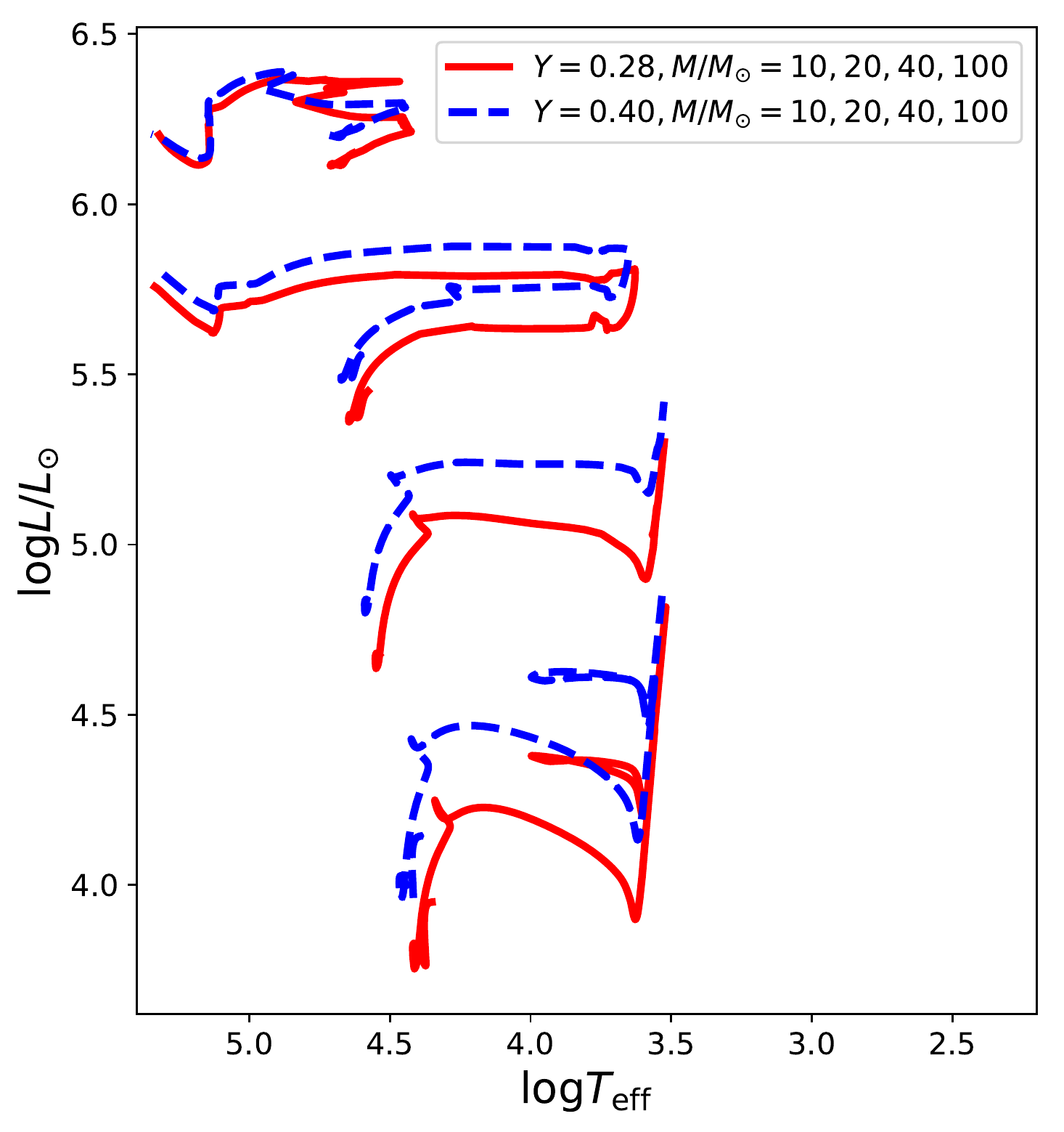}
\caption{Predicted evolution of high-mass stars on the Hertzprung-Russell diagram. The helium-enriched stars are systematically bluer and brighter, mimicking the effect of higher mass at fixed helium. At fixed initial mass, helium-enriched stars live substantially shorter lives.}
\label{fig:Tracks_LUTe}
\end{figure}

\begin{figure}
\centering
\includegraphics[width=0.45\textwidth, center]{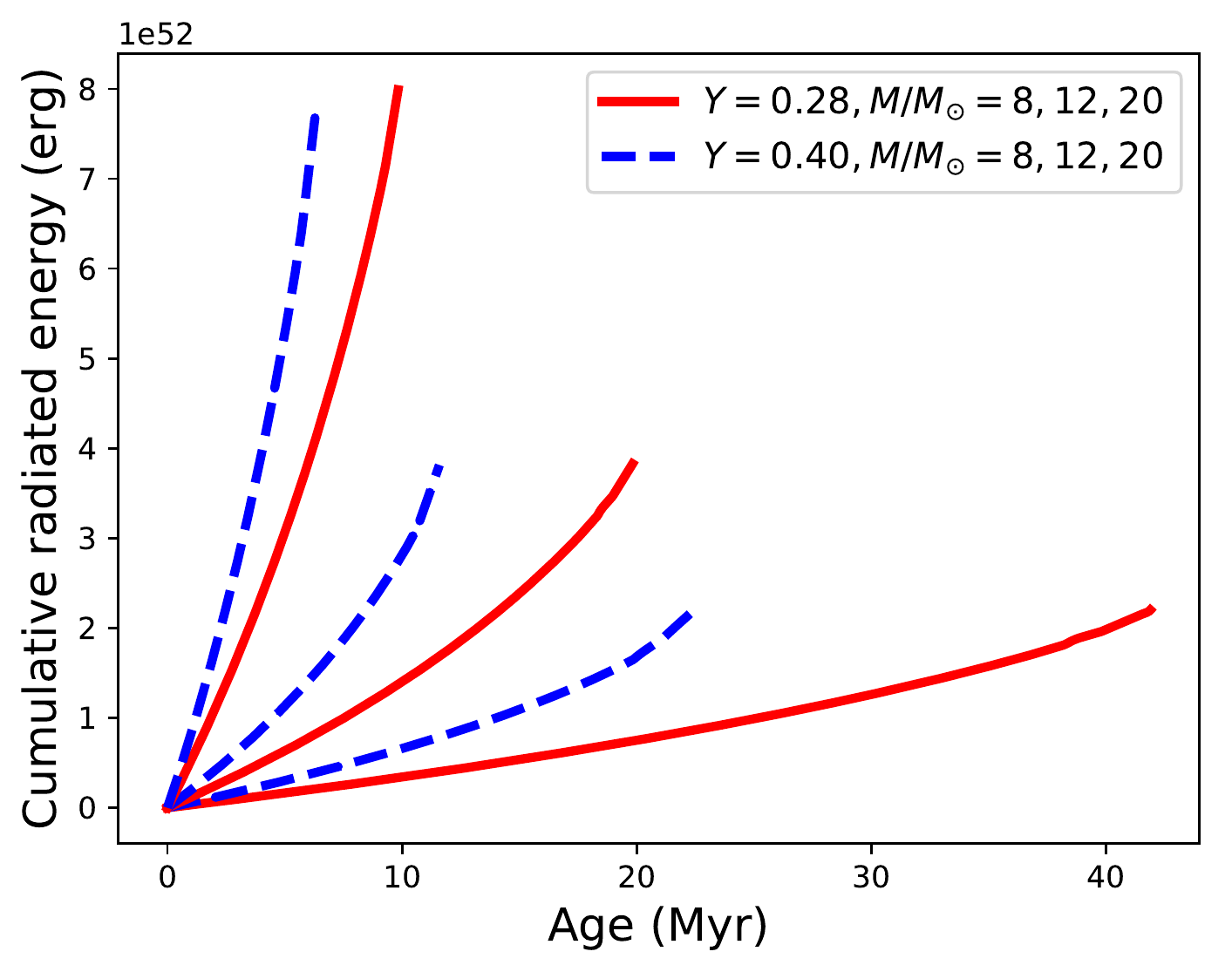}
\caption{The cumulative radiated energy of high-mass stars, as a function of mass and initial helium abundance, versus age. The helium-enriched stars achieve comparable total radiated energy at fixed mass, but do so with a shorter lifetime. As in Figure \ref{fig:Tracks_LUTe}, this is true for both the 20-100 $M/M_{\odot}$ tracks computed using the PARSEC v1.2 code.}
\label{fig:Tracks_Lage}
\end{figure}

We show, in Figure \ref{fig:Tracks_LUTe}, the predicted evolution of high-mass stars at both standard and enriched initial helium abundance on the  Herztprung-Russell diagram. In both cases, the more massive stars reach hotter temperatures, and are consistently more luminous at fixed temperature. That continues the trend seen in low-mass models discussed in the previous section. In Figure \ref{fig:Tracks_Lage} we show different information for the same stars -- the cumulative radiated energy as a function of age. The predicted energy released is negligibly-dependent on the initial helium abundance. For $M/M_{\odot}=8$ stars, the $Y=0.40$ star releases 6\% less total energy than the $Y=0.28$ star, and for $M/M_{\odot}=100$ stars, the reduction is 12\%. These shifts are smaller than the $\sim$20\% reduction in total available fuel. We have verified that the two sets of isochrones yield consistent answers in the part of parameter space where they both have predictions.

%\begin{figure}
%\centering
%\subfigure{\label{fig:HighMassTracks1}\includegraphics[width=70mm]{Figures/HighestMassStars.pdf}}
%\subfigure{\label{fig:HighMassTracks2}\includegraphics[width=70mm]{Figures/Tracks_LUTe.pdf}}
%\caption{Predicted evolution of high-mass stars on the Hertzprung-Russell diagram. The helium-enriched stars are systematically bluer and brighter, mimicking the effect of higher mass at fixed helium. That is true for both the 20-100 $M/M_{\odot}$ tracks computed using the PARSEC v1.2 code (TOP), and the 8-20 $M/M_{\odot}$ tracks from the work of \citealt{2009A&A...508..355B} (BOTTOM). }
%\label{fig:Tracks_LUTe}
%\end{figure}

%\begin{figure}
%\centering
%\subfigure{\label{fig:HighMassTracks3}\includegraphics[width=70mm]{Figures/Tracks_Lage_New.pdf}}
%\subfigure{\label{fig:HighMassTracks4}\includegraphics[width=70mm]{Figures/Tracks_Lage.pdf}}
%\caption{The cumulative radiated energy of high-mass stars, as a function of mass and initial helium abundance, versus age. The helium-enriched stars achieve comparable total radiated energy at fixed mass, but do so with a shorter lifetime. As in Figure \ref{fig:Tracks_LUTe}, this is true for both the 20-100 $M/M_{\odot}$ tracks computed using the PARSEC v1.2 code (TOP), and the 8-20 $M/M_{\odot}$ tracks from the work of \citealt{2009A&A...508..355B} (BOTTOM). }
%\label{fig:Tracks_Lage}
%\end{figure}

\begin{figure}
\centering
\includegraphics[width=0.44\textwidth, center]{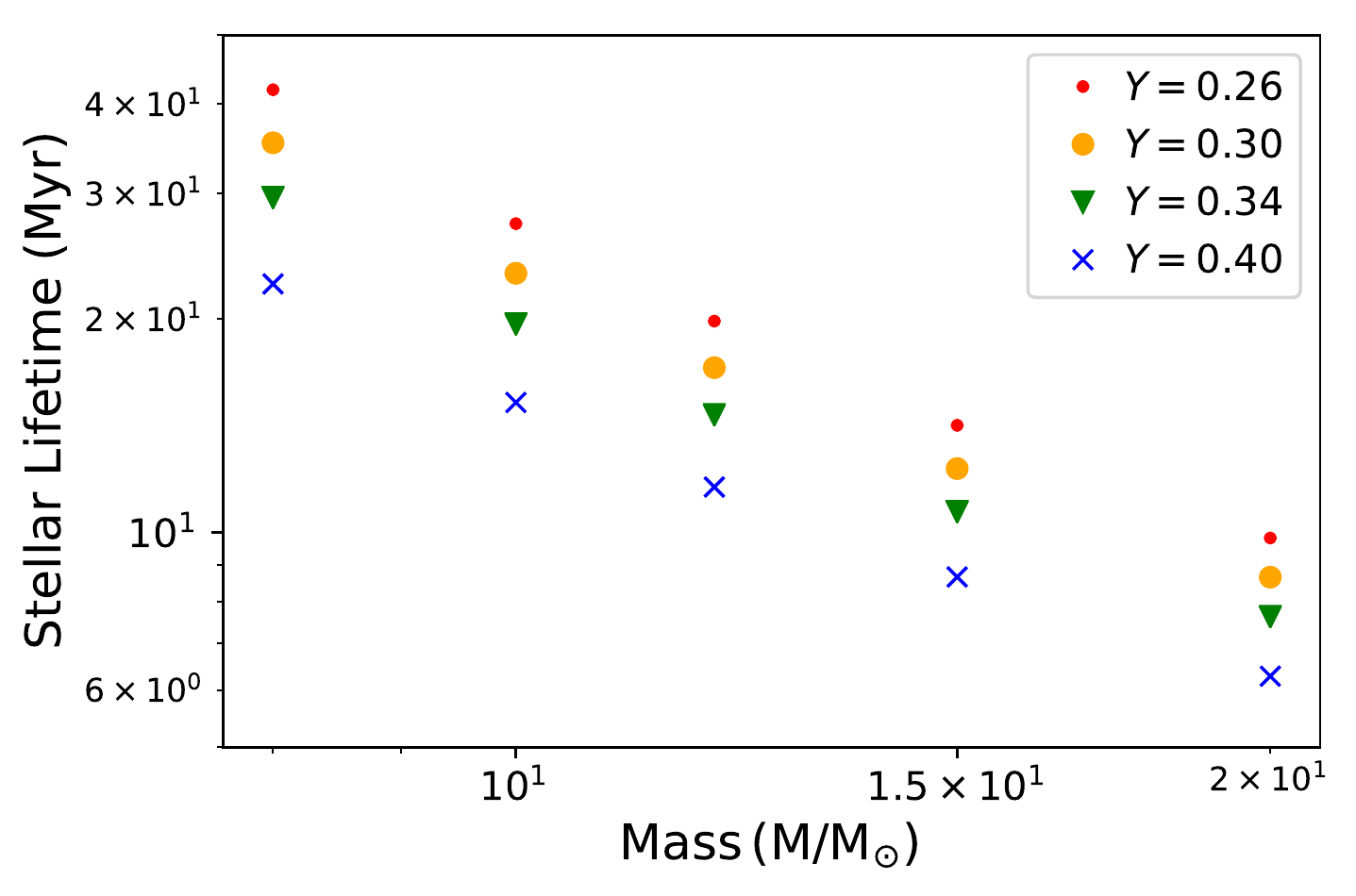}
\caption{The predicted stellar lifetimes are shown as a function of stellar mass and initial helium abundance.}
\label{fig:Tracks_Lifetimes}
\end{figure}

We show, in Figure \ref{fig:Tracks_Lifetimes}, the predicted lifetimes of high-mass stars as a function of mass and initial helium abundance. At fixed mass, stellar lifetime decreases with increasing initial helium abundance. That is due to three factors. The first is that helium-enriched stars are born with less hydrogen at fixed mass (by definition), and thus their main fuel source is already depleted. Second, the mean molecular weight is increased (as helium atoms are $\sim 4\times$ heavier than hydrogen atoms), and thus the interior equilibrium temperature and thus luminosity increases  as well. Finally, the reduction in the total number of free electrons per unit mass reduces the internal opacity of the star.

A multivariate least-squares fit of the stellar lifetime for tracks from \citet{2008A&A...484..815B,2009A&A...508..355B} satisfying $8 \leq (M/M_{\odot}) \leq 20$ and $0.26 \leq Y \leq 0.40$ yields the following scaling relation for the lifetime as a function of stellar mass:
\begin{equation}
    %\tau \approxprop M^{-1.49}{\times}\exp{[-3.82(Y-0.26)]}.
    \tau \approx (27\, \rm{Myr})\,\biggl(\frac{\mathit{M}}{10\,\mathit{M}_{\odot}}\biggl)^{-1.49}{\times}\exp{[-3.82(\mathit{Y}-0.26)]}.
    \label{EQ:Tracks_Lifetimes}
\end{equation}
The typical error in our coverage of this parameter space is less than 5\%. 

Equation (\ref{EQ:Tracks_Lifetimes}) is, unsurprisingly, not simply the continuation of fits done at the lower-mass regime \citep{2012AcA....62...33N,2014MNRAS.445..347K}. The effects of both mass and helium on the stellar lifetime are decreased at these higher stellar masses. Nevertheless, though the effect of helium enrichment on stellar lifetime is lessened relative to lower-mass stars, it is still present. For example, the lifetime of the $10 M_{\odot},\,Y=0.26$ star is 27 Myr, similar to that of the $7.5 M_{\odot},\,Y=0.40$ star.

If one assumes Equation (\ref{EQ:Tracks_Lifetimes}), an initial mass function of $dn/dm \propto m^{-2.35}$ for masses $M/M_{\odot} \leq 100$, and the same total stellar mass formed, the total amount of material that was in stars that have become remnants by t $= 27$ Myr will be $\sim$19\% higher for the $Y=0.40$ population than for the $Y=0.26$ population, and the total number of remnants will be $\sim$50\% higher. 

%The $10 M_{\odot}$ star with standard helium enrichment is predicted to live $\sim 27$ Myr (It's 27.2 Myr in the original stellar track). That roughly corresponds to the lifetime of the $\sim 7.5 M_{\odot}$ star at $y=0.40$. 

%In the next section, where we study the predicted integrated luminosities, this effect will be the dominant contributor: that by a given age the helium-enriched stellar population has burned through a larger fraction of its brightest stars. 

\subsection{The separate effects of helium and metallicity on high-mass stellar tracks}
In this investigation we are predominantly using tracks and isochrones of solar metallicity. %This metallicity need not be representative of the progenitors of globular clusters that will be observable at high redshift. 
Our results should be robust to this choice of metallicity as long as the predicted photometric observables of stellar populations have negligible cross-terms in their dependence on metallicity and initial helium abundance. 

We verify this in two different ways. We first compute the coefficients of Equation \ref{EQ:Tracks_Lifetimes} for a set of tracks shifted to lower metallicity by ${\Delta \log{Z}} = 1.23$ dex -- we find that the coefficients shift downward by $\sim$ 7\%. The same approximate non-covariance of the effects of metallicity and helium has already been demonstrated for predictions of lower-mass stars \citep{2014ApJ...784...32K}. We also similarly compute the dependence of the cumulative integrated luminosity -- energy -- released by stars as a function of stellar mass and initial helium abundance. Here, a shift to lower metallicity of ${\Delta \log{Z}} = 1.23$ dex results in shifts to the coefficients of less than 3\% or less. 

The assumption that the effects of metallicity and helium on predicted photometric observables of young stellar populations is non-covariant is thus sufficiently supported for this work. However, further study over a broader metallicity range will likely be warranted in the future.

\section{The predicted effect of enriched helium on the integrated light of starbursts}
\label{sec:HeliumPredictionStarbursts}

The previous section's discussion of high-mass, helium-enriched stellar tracks can facilitate our understanding of why helium enrichment might matter. However, we do not expect $JWST$ to observe \textit{isolated} massive helium-enriched stars, rather it will observe integrated light from populations of these stars.

%\textbf{We construct simple stellar populations from the tracks in a manner described by Shunsaku}, 

%We find that at fixed age and fixed stellar mass,  helium-enriched stellar populations are predicted to be brighter for nearly all ages and nearly all bandpasses, and cooler in nearly all differences of pairs of bandpasses. 

%\subsection{A note on the mass normalization of helium-enriched stellar populations}
%\label{subsec:Normalization}

In the following, we analyze the predicted effects of a globular cluster progenitor being helium-enriched at fixed mass, age, and metallicity. This is intended to isolate the effect of helium. In practice, helium-enriched stars are expected to have been formed in lesser numbers than helium-normal stars \citep{2008MNRAS.391..825D,2011A&A...533A.120V,2012ApJ...758...21C}, and the fraction of helium-enriched stars and the amplitude of their helium enrichment may be correlated with metallicity  \citep{2009A&A...505..117C,2017MNRAS.464.3636M}. Conversely, the helium-enriched populations might actually be slightly younger than the helium-normal populations, if they formed from the gas of the ejecta of the helium-normal stars, in which case they could temporarily dominate the flux budget of a globular cluster progenitor. We cannot account for that factor as, at this time, the age difference between helium-normal and helium-enriched populations remains uncertain.

\subsection{The predicted effect of enriched helium during the first 150 Myr}
\label{sec:First150Myr}

\begin{figure*}
\centering
\includegraphics[width=0.90\textwidth, center]{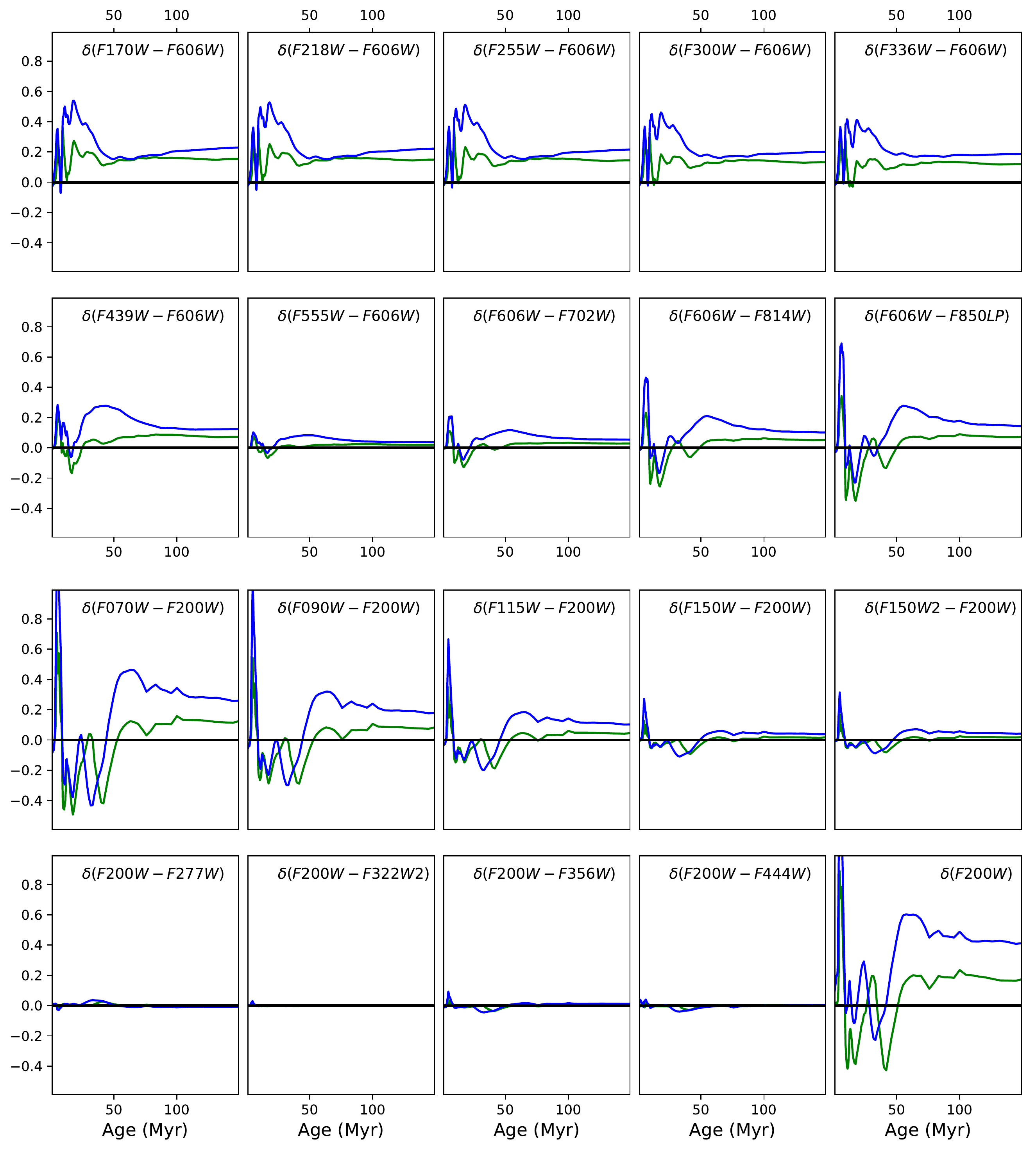}
\caption{Difference in predicted rest-frame colours for the helium-enriched ($Y=0.40$ in blue, $Y=0.33$ in green) relative to helium-normal ($Y=0.28$) populations as a function of age in the first 19 panels, with the difference in JWST-NIRCam $F200W$ magnitude shown in the bottom-right panel. A positive value denotes the helium-enriched population being more red or more bright in the given filter combination. }
\label{fig:AllMags}
\end{figure*}

\begin{figure*}
\centering
\subfigure{\includegraphics[width=154mm]{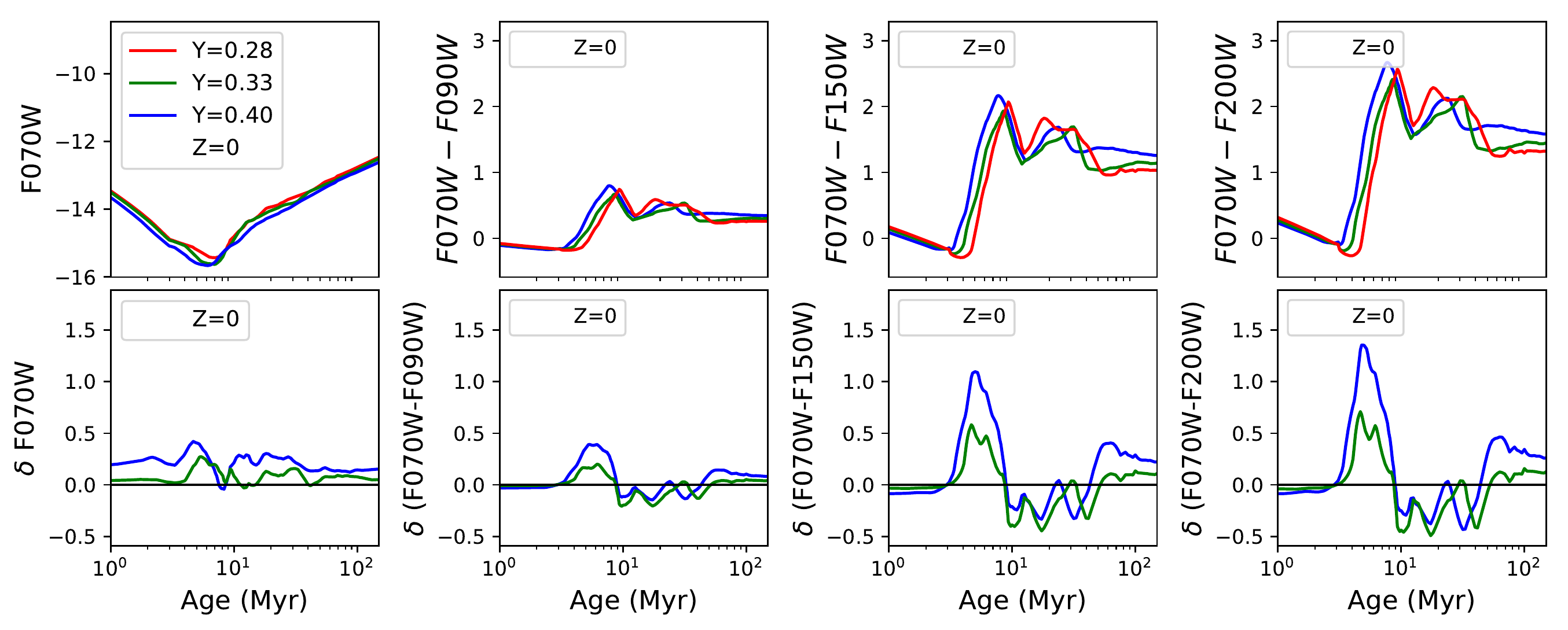}}\\[-7.7ex]
\subfigure{\includegraphics[width=150mm]{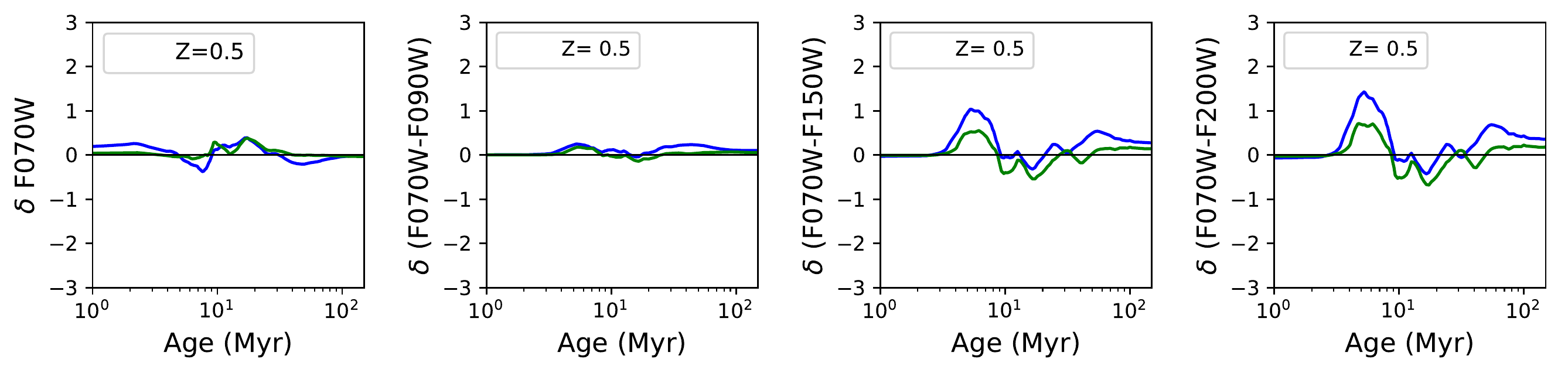}}\\[-7.7ex]
\subfigure{\includegraphics[width=150mm]{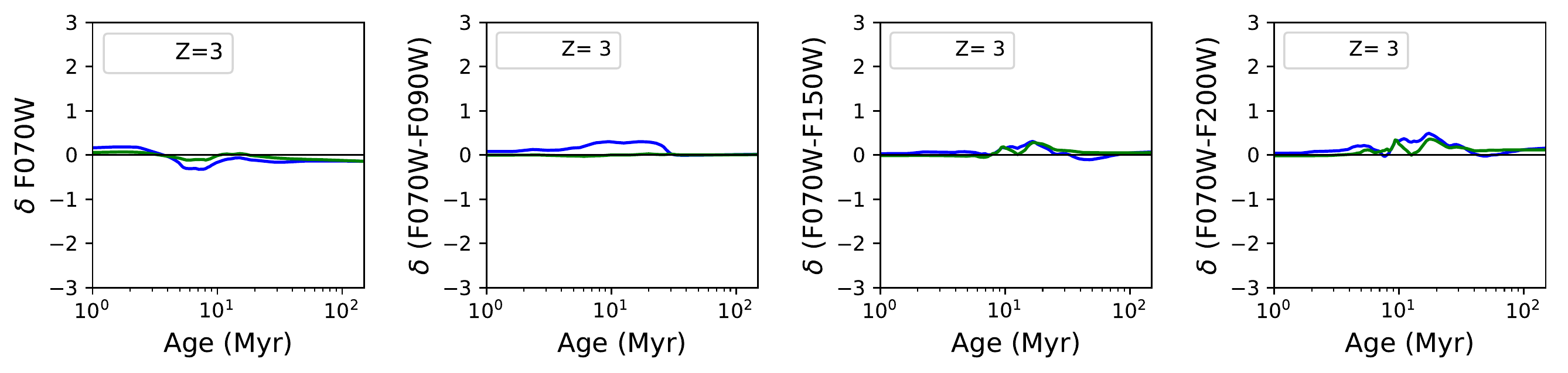}}\\[-7.7ex]
\subfigure{\includegraphics[width=150mm]{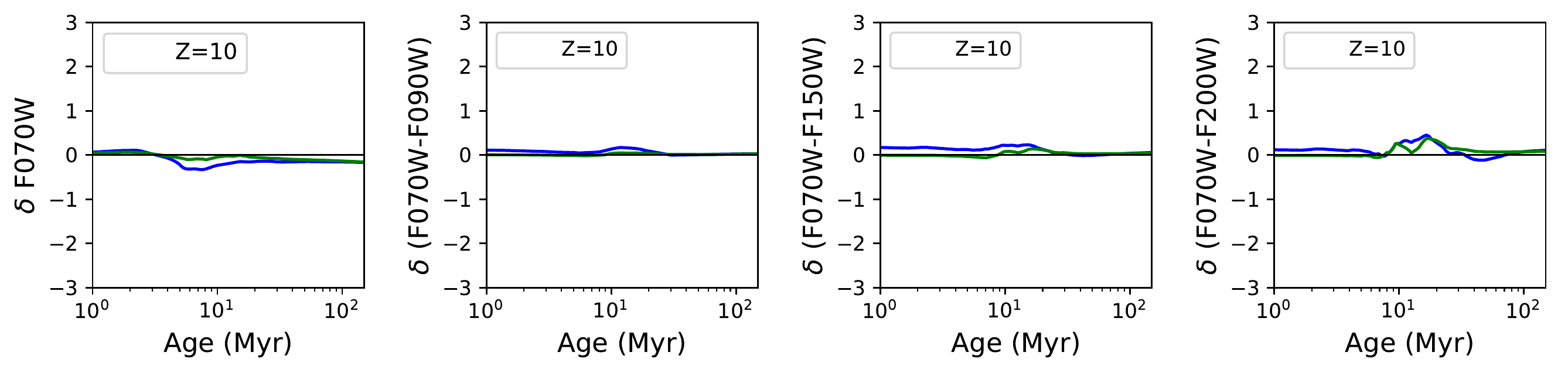}}
\caption{TOP: A sample of rest-frame magnitudes and colours for the helium-normal ($Y=0.28$, red), moderately helium-enriched ($Y=0.33$, green), and extremely helium-enriched synthetic populations ($Y=0.40$, blue). BOTTOM FOUR PANELS: The difference between the magnitude and colours of the  moderately helium-enriched ($Y=0.33$, green) and extremely helium-enriched synthetic populations ($Y=0.40$, blue) relative to that of the helium-normal population at four different cosmological redshifts. Here, positive values denote the helium-enriched population is brighter in the specified magnitude, or redder in the specified colour.  }
\label{fig:JWST_Mag_Colours}
\end{figure*}

We have computed the integrated magnitudes and colours of the helium-normal and helium-enriched populations for %four different \textit{HST} bandpasses at $z=0$ for heuristic value in Figure \ref{fig:HSTmag}, and 
for 12 \textit{HST} filters in the rest frame, and for 10 \textit{JWST}-NIRCam filters at 20 different redshifts. We show,  in Figure \ref{fig:AllMags}, 20 predicted differences in rest-frame predicted colours and magnitudes between the helium-normal and helium-enriched populations. We then show a selected subsample of predicted differences at various redshifts in Figure  \ref{fig:JWST_Mag_Colours}. There is no clear and simple statement that can be written as to the first-order difference between the helium-normal and helium-enriched populations, and what statements can be written are a function of age. We summarize the age-specific trends as follows:
\begin{itemize}
    \item In the first 5 to 10 Myr, the helium-enriched populations are predicted to be brighter at all wavelengths. That is due to their most massive stars more rapidly burning through their fuel. 
    \item There are several fluctuations in the period spanning 5 to 30 Myr after the starburst. The helium-enriched stars reach their red loops earlier, and are thus briefly much brighter. For example, at $F200W$ the $Y=0.40$ starburst is expected to be $\sim$1.6 mag brighter at $t \sim 5$ Myr, which then reverses as the most massive helium-enriched stars die earlier, and the helium-normal stars then also reach their red loops. 
    \item Fluctuations continue in the period spanning 30 to 50 Myr after the starburst, but the differences in magnitudes and colours are mostly small.
    \item From 50 Myr onward, the trends are more stable. The helium-enriched populations are fainter $\lambda \lesssim 5,000$ \AA, and brighter at longer wavelengths, with brightness offset increasing at longer wavelengths. The $Y=0.40$ population would converge to being $\sim 0.15$ brighter in $F070W$,  $\sim 0.30$ brighter in $F115W$,  and $\sim 0.40$ brighter in $F200W$. 
\end{itemize}

\begin{figure*}
\centering
\includegraphics[width=0.85\textwidth, center]{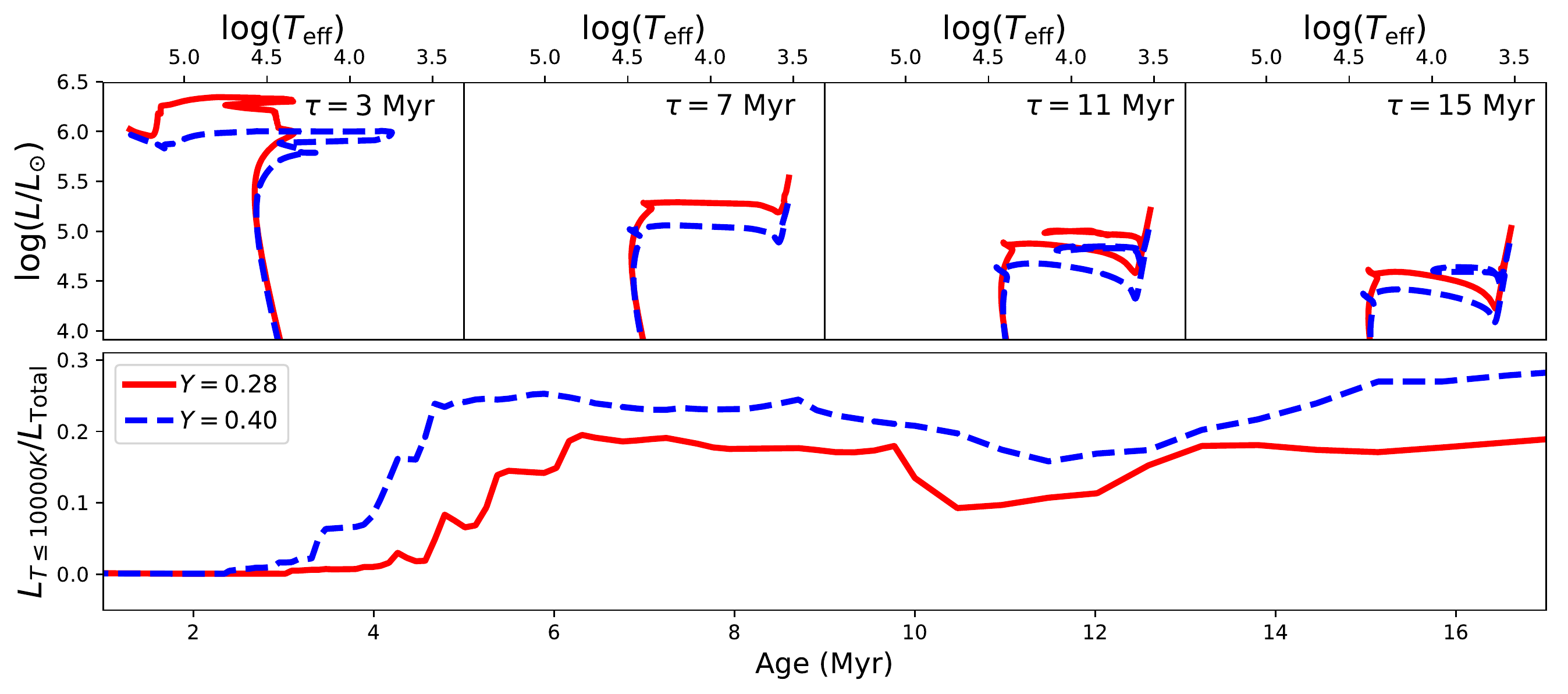}
\caption{The sudden emergence of red and blue loops cause rapid fluctuations in the predicted bolometric output of stellar populations, between the ages of 3 and 15 Myr. TOP: Isochrones for the helium-normal and helium-enriched populations for four representative ages. BOTTOM: Fraction of total luminosity from stars with $T_{\rm{eff}} \leq 10,000\,K$ as a function of age.}
\label{fig:IsochroneEvolution_2}
\end{figure*}

The origin of these rapid bolometric fluctuations for populations with $5 \lesssim \tau/\rm{Myr} \lesssim 30$ can be discerned from Figure \ref{fig:IsochroneEvolution_2}. We show, in the bottom panel, the fraction of the total luminosity predicted to come from stars colder than $T_{\rm{eff}} \leq 10,000\,K$ as a function of age. That fraction increases rapidly at $\tau \approx 4$ Myr, decreases again at $\tau \approx 12$ Myr, before then increasing and levelling off. The first increase corresponds to the emergence of a red extension to the Hertzprung-Russell diagram, which are present in the $\tau = 7$ Myr population, but not in that of the $\tau = 3$ Myr population. Similarly, there is a blue loop in the $\tau = 11$ Myr population, but not in the $\tau = 7$ Myr population. The effect of these rapid changes may increase the uncertainties in the eventual parameter estimates of globular cluster progenitors. In practice, our assumed age spread of ${\Delta}\tau = 3$ Myr may prove to be an underestimate for the progenitors of the most massive globular clusters. A larger ${\Delta}\tau$ would have the effect of reducing these fluctuations.

\begin{figure*}
\centering
\includegraphics[width=0.85\textwidth, center]{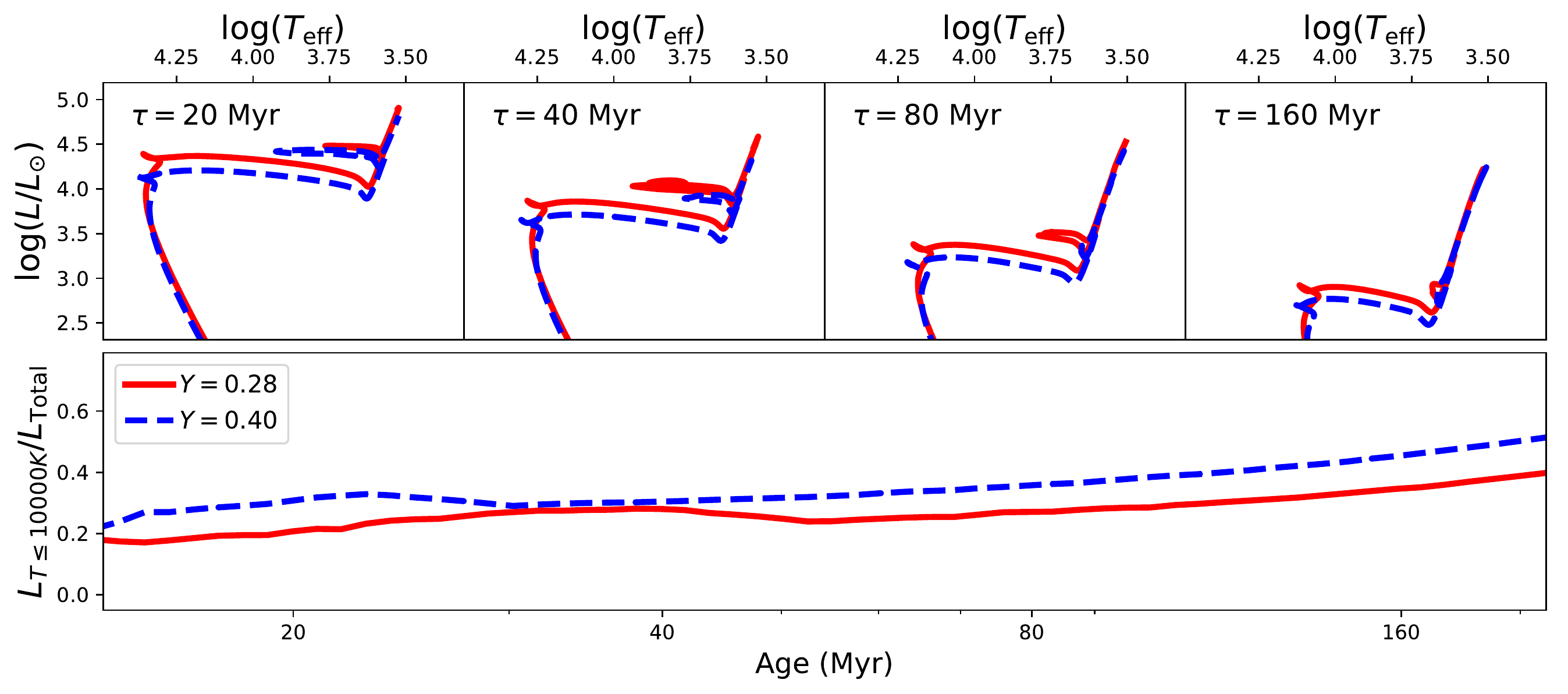}
\caption{Same as Figure \ref{fig:IsochroneEvolution_2}, but for later ages, with the ages in the bottom panel plotted logarithmically.  At these later ages, the differences between different stellar populations are a more stable function of time.}
\label{fig:IsochroneEvolution_3}
\end{figure*}

We show, in Figure \ref{fig:IsochroneEvolution_3}, how the differences between the helium-normal and helium-enriched populations continue through to $\tau=160\,$Myr. There are still some variations, but they are smaller than they were for ages $\tau \lesssim 30\,$Myr. The helium-enriched population, to first order, continuously gets an additional 10\% of its integrated luminosity from stars with $T_{\rm{eff}} \leq 10,000\,K$. 

That the helium-enriched populations converge to being cooler, and thus redder in most filter combinations, may be counterintuitive, as the general literature on helium-enriched models, discussed in Section \ref{sec:StellarModels}, often refers to how the tracks of individual helium-enriched stars are generally hotter than those of helium-normal stars at fixed initial mass and evaluated at the same evolutionary phase,  with this shown again in Figure \ref{fig:Tracks_LUTe}. The shift to redder integrated colours occurs due to the fact that helium-enriched populations evolve faster, and thus they will contain proportionately more red giants, at earlier times. 

The time-dependent and redshift-dependent predictions for globular cluster progenitors in the \textit{JWST} bandpasses are listed in Table \ref{table:magredshiftcomparison}.

\begin{figure}
\centering
\includegraphics[width=0.45\textwidth, center]{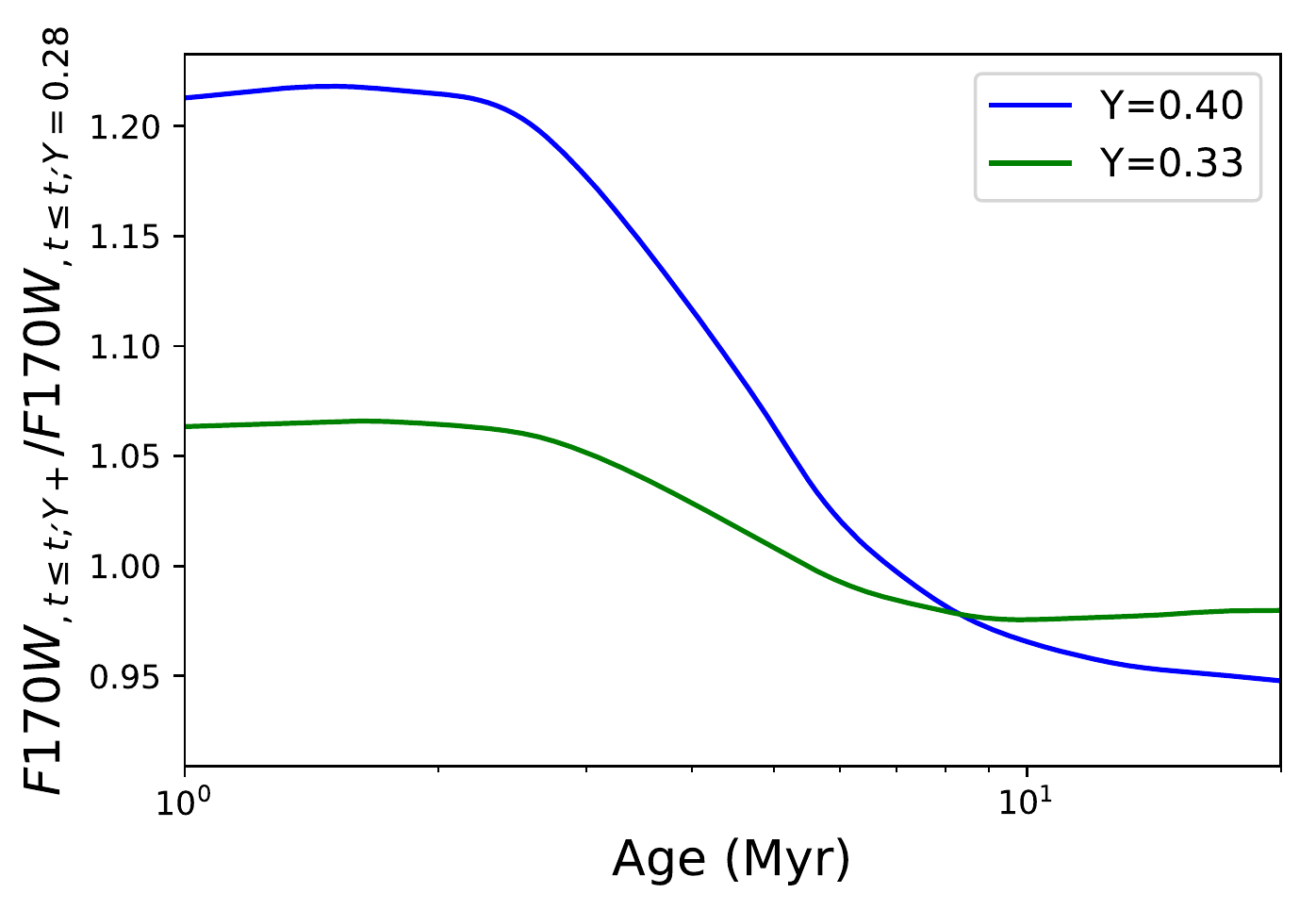}
\caption{The predicted cumulative ratio of flux in the $F170W$ filter, which we use as a proxy for ionizing flux, as a function of time, for the $Y=0.40$ and $Y=0.33$ populations relative to the $Y=0.28$ population. The predicted cumulative ionizing flux from the helium-enriched populations is similar to that of the helium-normal population. }
\label{fig:F170ratio}
\end{figure}

We use the cumulative, integrated flux for the $F170W$ bandpass as a proxy for the ionizing radiation, which we plot in Figure \ref{fig:F170ratio}. The helium-enriched populations are predicted to emit more of these (nearly)-ionizing photons -- but only for the first $\sim$5 Myr. Subsequent to this, the cumulative radiation at these wavelengths is expected to be slightly ($\lesssim 5\%$) lower. 

%\begin{figure*}
%\centering
%\includegraphics[width=1.00\textwidth, center]{Figures/BolometricMagnitudes.pdf}
%\caption{For fixed initial stellar mass, the difference in magnitude between the $Y=0.40$ population (blue) and $Y=0.33$ population (green) with that of the $Y=0.28$ population. We plot the variable  $\delta{X} = X_{Y=0.28}-X_{Y=0.40}$ as a function of time, so positive values mean that the helium-enriched population will be brighter in that bandpass.}
%\label{fig:BolometricMassive}
%\end{figure*}

%\begin{figure*}
%\centering
%\includegraphics[width=1.00\textwidth, center]{Figures/KeyMagsColours.pdf}
%\caption{TOP: A few predicted integrated magnitudes and colours for the first 20 Myr of a stellar population. BOTTOM: The differences in colours and magnitudes, where $\delta{X} = X_{Y=0.28}-X_{Y=0.40}$. The helium-enriched populations are generally predicted to be slightly brighter and redder, but there is some noise on that trend due to the mass-loss schedule among the highest-mass stars, which is also uncertain. }
%\label{fig:KeyMagColours}
%\end{figure*}

%\begin{figure*}
%\centering
%\includegraphics[width=150mm]{Figures/BolometricMagnitudes_HST.pdf}
%\caption{The difference between four selected \textit{HST} WFC3/UVIS magnitudes of the  moderately helium-enriched ($Y=0.33$, green) and extremely helium-enriched synthetic populations ($Y=0.40$, blue) relative to that of the helium-normal population. Here, positive values denote the helium-enriched population is brighter in the specified magnitude.}
%\label{fig:HSTmag}
%\end{figure*}

\begin{table}
\begin{center}
\begin{tabular}{|cc| ccc  |}
\hline
log(t/Myr) & $z$ & ${\Delta}F070W$ &  ${\Delta}F090W$ &  ${\Delta}F200W$ \\ 
\hline\hline
-1.00 & 0.0 & 0.16 & 0.13 & 0.07\\ 
-0.60 & 0.0 & 0.17 & 0.13 & 0.07\\ 
-0.20 & 0.0 & 0.18 & 0.15 & 0.09\\ 
0.20 & 0.0 & 0.22 & 0.20 & 0.16\\ 
0.60 & 0.0 & 0.30 & 0.45 & 1.01\\ 
1.00 & 0.0 & 0.21 & 0.10 & -0.04\\ 
1.40 & 0.0 & 0.25 & 0.26 & 0.23\\ 
1.80 & 0.0 & 0.14 & 0.28 & 0.60\\ 
-1.00 & 0.3 & 0.16 & 0.17 & 0.07\\ 
-0.60 & 0.3 & 0.16 & 0.17 & 0.07\\ 
-0.20 & 0.3 & 0.17 & 0.18 & 0.09\\ 
0.20 & 0.3 & 0.22 & 0.23 & 0.14\\ 
0.60 & 0.3 & 0.14 & 0.29 & 0.89\\ 
1.00 & 0.3 & 0.18 & 0.22 & -0.00\\ 
1.40 & 0.3 & 0.12 & 0.25 & 0.25\\ 
1.80 & 0.3 & -0.06 & 0.13 & 0.55\\ 
-1.00 & 10.0 & 0.48 & 0.32 & 0.16\\ 
-0.60 & 10.0 & 0.48 & 0.32 & 0.16\\ 
-0.20 & 10.0 & 0.50 & 0.33 & 0.16\\ 
0.20 & 10.0 & 0.42 & 0.33 & 0.20\\ 
0.60 & 10.0 & 0.06 & 0.14 & -0.01\\ 
1.00 & 10.0 & 0.86 & 0.72 & -0.10\\ 
1.40 & 10.0 & 2.73 & 0.73 & -0.10\\ 
1.80 & 10.0 & -0.14 & -0.25 & -0.13\\ 
\hline
\end{tabular}
\caption{A subsample of predicted magnitude differences between a helium-enriched ($Y=0.40)$ and helium-normal population ($Y=0.28$) as a function of population age and cosmological redshift $z$. Here, a positive value of ${\Delta}$mag means that the helium-enriched population is brighter. A version of this table which has denser sampling in age and redshift and covers 10 \textit{JWST} bandpasses is available in the online addition. }
\label{table:magredshiftcomparison}
\end{center}
\end{table}

\section{Other possible systematics with young, helium-enriched stellar populations}
\label{sec:OtherEffects}

\subsection{The effect of elevated helium abundance on the model atmospheres of stars}
\label{sec:StellarAtmospheres}

The prediction of ionizing radiation from the preceding section does not trivially follow from stellar theory. As discussed previously, a shift in $Y$ leads to shifts in $T_{\rm{eff}}$, and $\log{g}$ at fixed age, metallicity, and stellar mass. The stellar models used in this work have computed this shift, and have also assumed a mapping, from model atmospheres and synthetic spectra, to convert atmospheric parameters to various bandpass-dependent fluxes. However, this mapping might itself depend on $Y$, if the atmospheric structure depends on $Y$. We explore this issue here. 

%What stellar models typically predict are bolometric luminosity, effective temperature, and surface gravity. A combination of model atmospheres and bolometric corrections are then used to convert these into wavelength-dependent flux distributions, and eventually, filter-specific magnitudes. However, model atmospheres and bolometric corrections have not been tested (and to our knowledge, investigated), in the helium-enriched O-star regime. 

It is conceivable that the increased helium abundance could modify the atmospheric opacity of hotter stars. Among other changes, the decreased abundance of hydrogen necessitates a decreased stellar opacity to radiation that could contribute to the re-ionization of the universe. 

%There are two reasons for this. First, helium-enriched atmospheres should have fewer free electrons. Even if all of the hydrogen and helium are ionized, helium atoms have half as many electrons per unit mass. Second, the increased abundance of helium maps exactly onto a decreased abundance of hydrogen, and thus potentially a noticeable decrease in Lyman-$\alpha$ opacity. 

We thus computed synthetic spectra for several cases. We first discuss the following four: ($\log{T_{\rm{eff}}}$, $\log{g}$) = (4.3899 3.8718) and  (4.4716 3.9959), for $Y=0.26,0.40$. We use the models of \citet{1970SAOSR.309.....K,1981SAOSR.391.....K,1993KurCD..18.....K,2005MSAIS...8...14K,2013ascl.soft03024K}. Each model atmosphere assumes $\log(Z/Z_{\odot})=-1.23$ on the abundance scale of \citet{2009ARA&A..47..481A}. The temperature and gravity values are taken from the helium-normal and helium-enriched tracks. Without correcting for atmospheric effects, the helium-enriched stars will automatically emit more ionizing radiation as they reach higher surface temperatures (Figure \ref{fig:Tracks_LUTe}). We find that the models predict a $\sim$1\% increase in total flux in the 1,000 \AA $\leq \lambda \leq$ 1,500 \AA\, range.

We also compute model atmospheres at ($\log{T_{\rm{eff}}}$, $\log{g}$) = (4.3899 3.8718) with both an $\alpha$-enhanced, and a CNONa-extreme mixture, which is expected for the most helium-enriched globular cluster populations. Both mixtures have shifts of $+0.40$ dex in the $\alpha$-abundances, and the second mixture has shifts of $-0.60,+1.80,-0.80,+0.80$ dex in the respective abundances of CNONa. The elemental abundance mixtures are respectively taken from \citet{2006ApJ...642..797P} and \citet{2009ApJ...697..275P}.  We find that the models predict a $\sim$0.1\% increase in total flux in the 1,000 \AA $\leq \lambda \leq$ 1,500 \AA\, range for the CNONa mixture, relative to the $\alpha$-enhanced mixture. These effects are thus predicted to be negligible. 

We note that the estimates of this subsection apply only to massive main-sequence stars that still have their birth composition, and that have surface temperatures colder than 30,000 Kelvin. Readers interested in the effects at higher surface temperatures, and for massive stars with even more extreme CNO and He abundances due to evolution from factors such as internal mixing, are referred to \citet{2020MNRAS.494.3861R}.

%The models predict negligible difference in the ultraviolet flux. The difference in flux for both the lower and higher temperature case is $\sim$1\% or less in the range 1000$\leq\lambda$/{\AA} $\leq$5000. In the lower temperature case, the ratio of flux rises from 1.006 to 1.016 as \AA increases from 1500\AA to 5000\AA.  In the higher temperature case, the ratios are even smaller, 0.997 and 1.009.  

\subsection{The plausibility of a higher Lyman-$\alpha$ escape fraction}
\label{subsec:EscapeFraction}
 The recent work of \citet{2018MNRAS.479..332B} assumes that the escape fraction of ionizing radiation from globular clusters was $f_{\rm{esc}}(t < t_{\rm{SNeII}}) = 0$ and $f_{\rm{esc}}(t > t_{\rm{SNeII}}) = 1$. That is consistent with earlier work arguing for an escape fraction close to unity \citep{2002MNRAS.336L..33R,2004ASPC..322..509R}. The assumption of a high escape fraction after the supernovae explosions is not simply the heuristic argument that the shock of the explosions should drive out the gas. It is also due to measurements showing that second-generations of globular clusters have little or no iron enrichment \citep{2009A&A...508..695C}. Even in the case of the metal-diverse globular cluster $\omega$ Cen (NGC 5139), it can be estimated that 99.8\% of the supernovae ejecta were lost from the cluster \citep{2013MmSAI..84..162R}. That estimate arises from the fact that though some stars in $\omega$ Cen have higher metallicities, they are not sufficiently high to account for all of the core-collapse supernovae of the more metal-poor stars in that cluster.  This demonstrates that the supernovae ejecta were not held on to, and thus makes it unlikely that the remaining initial gas was held on to as well, as the two should be mixed at some level. 
 
  A prediction that is at odds with the assumptions of \citet{2018MNRAS.479..332B} is that from the detailed simulations of \citet{2015MNRAS.451.2544P}. The middle panel of their figure 5 shows the distribution of escape fraction as a function of stellar mass and redshift. The typical escape fraction of ionizing photons for systems with $M/M_{\odot} = 10^6$ is $f_{\rm{esc}} \approx 0.01$. We note that the scatter is large in all displayed bins of redshift and star formation history, though the prediction of $f_{\rm{esc}}$ is that it will virtually always be lower than 10\% for $M/M_{\odot} \lesssim 10^7$.
  
  Another result of \citet{2015MNRAS.451.2544P} that is of interest to this study is that shown in their Figure 3. The escape fraction of ionizing radiation is (unsurprisingly) predicted to be a decreasing function of increasing hydrogen column density $N_{H}$, for all redshifts and all star formation histories. A $\sim$10,000-fold increase in $N_{H}$ results in a $\sim$100-fold decrease in $f_{\rm{esc}}$ (with substantial scatter). 
  
  This trend may be relevant for helium-enriched second generations, if their surrounding interstellar medium is also enriched in helium and correspondingly deficient in hydrogen. The opacity to Lyman-$\alpha$ photons would be reduced, since helium is effectively transparent to nearly all stellar radiation. Helium is not just a noble gas, but the noblest of them all. The first ionizing potential of helium is 24.62 eV, the highest of all of the elements, and some $\sim$80\% higher than that of hydrogen. The Lyman-$\alpha$ escape fraction of second generation globular cluster starbursts should thus be increased. 
  
  This argument assumes that the efficiency of converting gas into stars does not change as the helium-to-hydrogen ratio changes. This may be incorrect - \citet{2012ApJ...757..132H} argue in their Section 4.2.3 that gas enriched in He, N, Na and depleted H, C, O, and Ne, as found in extreme globular cluster populations, would have a radiative cooling coefficient that is up to 0.50 dex higher than gas with a canonical composition, at fixed total metallicity. That is approximately equivalent to an increase of 1 dex in the total  metallicity of the gas. We ignore this caveat here, and thus our estimate of increased ionizing flux is an underestimate, as the gas may be depleted in total mass in addition to being depleted in hydrogen. 

Given that the number of hydrogen atoms is proportional to the mass in hydrogen of the star-forming cloud, the relationship between the optical depth of the first ($ \tau_{\rm{GenI}}$) and of the second ($\tau_{\rm{GenII}}$) stellar generation should go as:
\begin{equation}
    \tau_{\rm{GenII}} \approx  \biggl( \frac{1-Y_{2}}{0.75} \biggl) \tau_{\rm{GenI}},
\end{equation}
where $Y_{2}$ is the helium mass fraction of the second generation.  The helium mass fraction inferred in the most massive globular clusters, $Y=0.40$, would thus result in a fractional decrease of the optical depth of $(1-0.6/0.75) = 1-0.80 \rightarrow 20$\%. 

\begin{figure}
\centering
\includegraphics[width=0.40\textwidth, center]{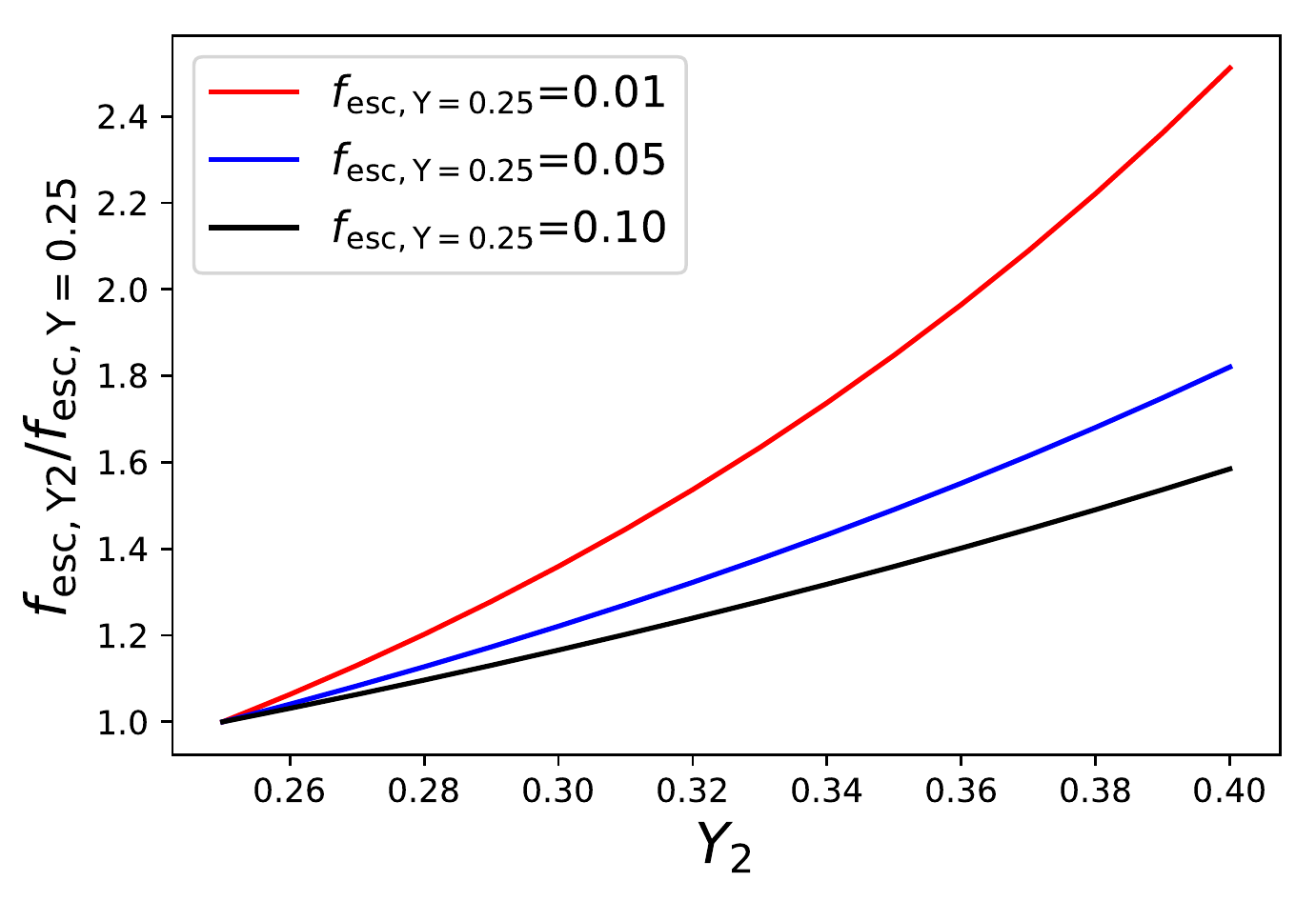}
\caption{The fractional increase of the escape fraction of the second generation relative to the first generation, ${f_{\rm{esc,GenII}}}/{f_{\rm{esc,GenI}}}$, as a function of the helium-mass fraction of the second generation ($Y_{2}$), and the escape fraction of the first generation ($f_{\rm{esc,GenI}}$).}
\label{fig:AnalyticEscapeFraction}
\end{figure}

The definition relating the optical depth to the escape fraction $f_{\rm{esc}}$ is:
\begin{equation}
     e^{-\tau} \equiv f_{\rm{esc}}.
\end{equation}
It follows that a 1\% escape fraction for the first generation will translate to one as high as 2.5\% for the second generation, and that a 10\% escape fraction for the first generation will translate to one as high as 16\% for the second generation. Indeed, the fractional increase of the escape fraction goes as:
\begin{equation}
    \frac{f_{\rm{esc,GenII}}}{f_{\rm{esc,GenI}}} =
     \exp \left( \biggl[ \frac{Y_{2}-0.25}{0.75} \biggl] \log{f_{\rm{esc,GenI}}} \right)  
     \label{EQ:escapefraction}
    %e^{-\biggl(  }
\end{equation}
We show the fractional increase in escape fraction as a function of the initial escape fraction $f_{\rm{esc,GenI}}$ and the helium mass fraction of the second generation in Figure \ref{fig:AnalyticEscapeFraction}.

It thus appears unlikely that the peculiar physics of helium-enriched stellar populations are a substantial perturbation on the reionization budget. Their escape fraction will only be substantially increased in the regime of high opacity, where there are few photons escaping in any case. However, prior calculations of $f_{\rm{esc}}$ for globular cluster progenitors predict values close to unity \citep{2002MNRAS.336L..33R,2004ASPC..322..509R}, and thus the change would be negligible. It was also shown in the previous section that their cumulative ionizing flux will only be marginally increased.

\subsection{Variations in initial helium abundance and the predicted final core masses of massive stars}
\label{sec:FinalCoreMasses}

\begin{figure}
\centering
\includegraphics[width=0.40\textwidth, center]{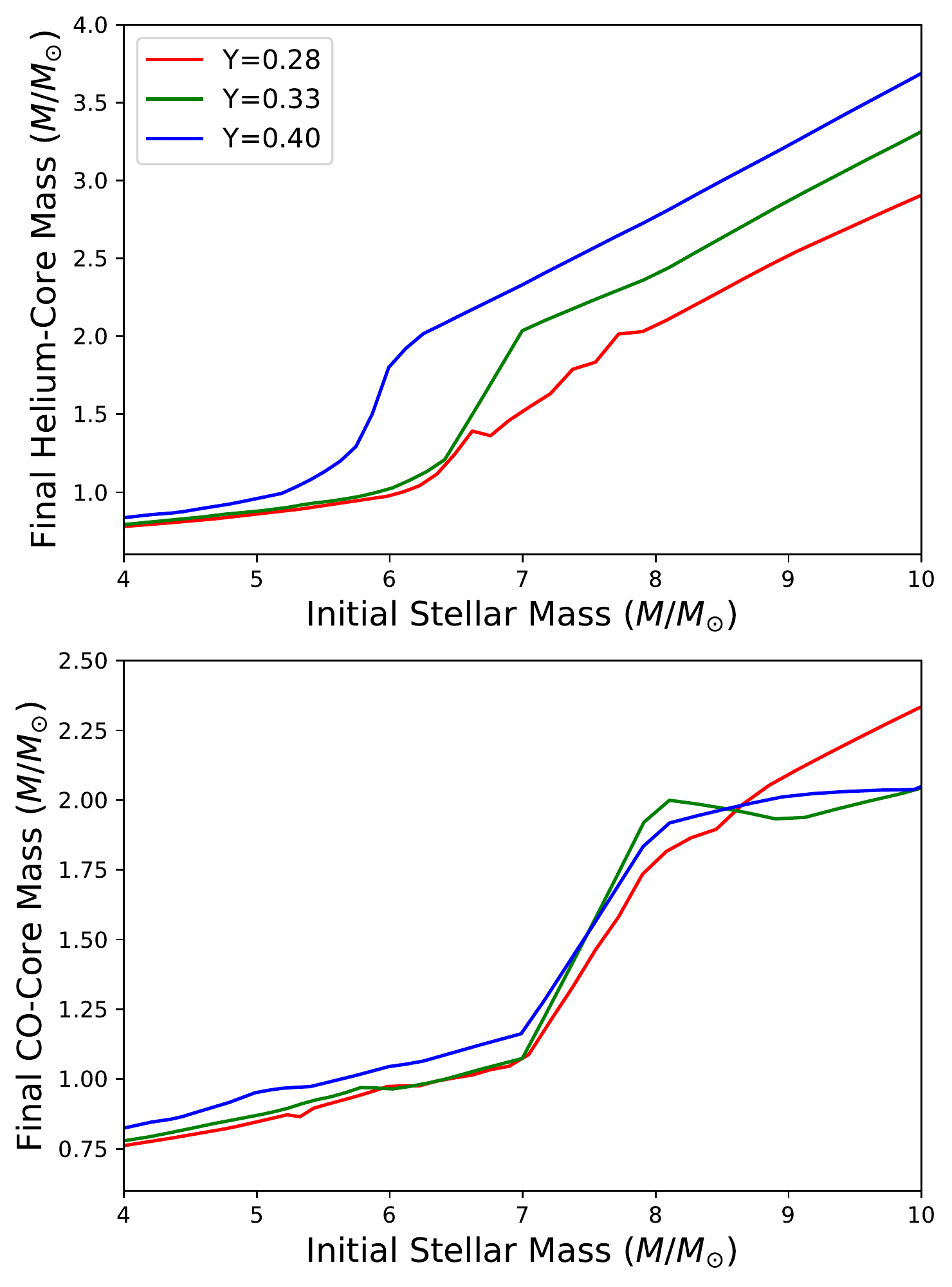}
\caption{The final masses of Helium-core (top panel) and carbon-oxygen core (bottom panel) as a function of initial stellar mass and initial helium abundance, as predicted by the PARSEC v1.2 isochrones.}
\label{fig:CoreMasses}
\end{figure}

We show, in Figure \ref{fig:CoreMasses}, the predicted final helium core masses and carbon-oxygen core masses as a function of initial mass and varying initial helium abundance. These are the values predicted by the PARSEC v1.2 isochrones. 

The predictions are for a substantial increase in the final helium-core mass, and a modest increase in the final carbon-oxygen core mass, as the initial helium abundance is increased. This continues the trend, predicted for lower mass stars, of larger mass remnants as a function of initial mass for helium-enriched populations \citep{2014ApJ...784...32K,2017A&A...602A..13C}. 

Our findings are thus qualitatively consistent with the argument of \citet{2017MmSAI..88..244K}, that helium enrichment could shift the mass threshold for core-collapse supernovae at fixed metallicity. We cannot ascertain quantitative consistency at this time as this would require a separate investigation. However, this has the potential to contribute an offset to the integral of ionizing flux, as the supernovae themselves contribute a lot of ionizing radiation, and more massive cores at fixed initial mass and fixed initial mass function would necessitate a greater number of core-collapse supernovae.

\section{Discussion and Conclusion}
\label{sec:Conclusion}

In this work, we have investigated the predicted effects of higher initial helium abundances in the flux of massive globular cluster starbursts. We have done this using two sets of tracks and isochrones. We predominantly made use of tracks and isochrones which were computed for this work using the PARSEC v1.2 code \citep{2012MNRAS.427..127B,2014MNRAS.445.4287T,2015MNRAS.452.1068C}. We also used the tracks and isochrones from the works of \citet{2008A&A...484..815B}
and \citet{2009A&A...508..355B} as a consistency check. 

We found that at fixed metallicity, age, and total stellar mass, helium-enriched stellar populations are expected to be redder from $\sim$50 Myr onward, with rapid fluctuations taking place before that time. The helium-enriched $({\Delta}Y=0.12)$ population then converges to being fainter for $\lambda \lesssim 0.50\,{\mu}m$, and up to 0.40 mag brighter at $\lambda \approx 2.0\,{\mu}m$. 

Given that the first set of observations of globular cluster progenitors will likely be very coarse, we do not expect the effect of helium abundance variations to be immediately disentangled. A likelier outcome is that, if a helium-enriched population is observed, it will be misconstrued as having the wrong age, metallicity, and mass. However, it is also the case that there will often be no satisfactory fit if observations are made in a sufficient number of bandpasses, which may result in an interpretation of helium enrichment. 

In Section \ref{sec:OtherEffects}, we explored some other possible effects. We found that the amount of radiation capable of ionizing hydrogen could be shifted higher, but only marginally so, due to two effects. First, the O-stars should themselves have a slightly lower opacity to ionizing radiation, but this effect turned out to be very small. Second, the interstellar medium surrounding these globular clusters might itself be depleted in hydrogen due to being enhanced in helium, and thus the escape fraction would be shifted higher. A potentially larger effect is that of the higher final core masses predicted by the PARSEC v1.2 models, which would suggest more massive remnants, and a potentially greater number of core-collapse supernovae. 

There are several additional sources of systematic uncertainty that we have not accounted for. We discuss four of these:
\begin{itemize}
    \item We assumed that helium-normal and helium-enriched stellar populations would form massive stars with the same initial mass function. This assumption can be a functional hypothesis, but it is without empirical support. However, the IMF is believed to be primarily a result of turbulent fragmentation, moderated by magnetic fields and stellar feedback from jets and radiation \citep{2002ApJ...576..870P,2009ApJ...702.1428H,2012MNRAS.423.2037H,2014ApJ...790..128F,2019FrASS...6....7K} and is therefore unlikely to depend on the effects of helium enrichment discussed here.
    \item Similarly for the binary fraction, which could create more (or fewer) massive stars and altered evolutionary channels by means of mergers and mass transfers. There is  evidence that the lower-mass stars of the second generations of globular clusters have lower binary fractions \citep{2010ApJ...719L.213D,2015A&A...584A..52L}, but the implications for the binary fraction at higher masses are uncertain.
    \item Similarly for rotation, which we neglected. We did not explore whether rotation would have a different evolutionary effect on helium-enriched stars than it does on helium-normal stars, nor whether we should expect the distribution of rotation rates to be a function of helium abundance. Variations in rotation would have an effect, as increased rotation results in increased luminosity at fixed metallicity and mass for massive stars  \citep{2011A&A...530A.115B}. Rotation also has an effect on the observed $T_{\rm{eff}}$, magnitudes and colours of massive stars in a manner that depends on the inclination angle between the star's rotation axis and the line of sight to the star \citep{2019MNRAS.488..696G}. 
    There is evidence for a rotational dichotomy in intermediate-age clusters in the Large Magellanic Cloud \citep{2018ApJ...864L...3G,2018MNRAS.480.3739B}, but it is not known if this would have also been the case in the massive stars of globular clusters, and how it would have mapped onto populations with different abundances. 
    \item Mass loss will have a substantial effect. This can be easily discerned by, for example, inspecting Figures \ref{fig:IsochroneEvolution_2}  and seeing the magnitude variations in the 0--20 Myr range. This will undoubtedly challenge attempts at interpreting $JWST$ data -- there are no empirical constraints on mass loss for massive stars with [Fe/H] $\leq -1.0$, as there are no such stars in the local group. Further, the assumption that mass loss is similar for massive helium-normal and helium-enriched stars may not be correct -- it is seemingly incorrect at low masses. The relative paucity of abundance anomalies on the asymptotic giant branches of globular clusters \citep{1981ApJ...244..205N,2013Natur.498..198C,2016MNRAS.460L..69M} suggests that mass-loss may be increased in helium-enriched stars, causing those stars to bypass the asymptotic giant branch and go straight to the hot horizontal branch stars. Those are the AGB-manqu\'e stars, originally postulated by \citet{1990ApJ...364...35G}, to explain the excess ultraviolet luminosity of metal-rich elliptical galaxies \citep{1983HiA.....6..165F,1988ApJ...328..440B,2018ApJ...857...16G}.
\end{itemize}
 
 There is thus a need for further theoretical work in the coming years, as observations of young, massive star clusters at high-redshift come in. These will augment the current samples (from lensed systems, \citealt{2017arXiv171102090B,2019arXiv190407941V}), by at least two orders of magnitude \citep{2019MNRAS.485.5861P}. Characterizing such systems, quantifying their contribution to reionization, and relating them to the populations of old globular clusters seen today may become one of the major emerging fields of astronomy.

\section*{Acknowledgments}
We thank the referee for a diligent and constructive referee report that greatly improved the manuscript.

We thank Todd Thompson, Sadegh Khochfar, S. Michael Fall, Anna Lisa Varri, Ivan Cabrera-Ziri, Leo Girardi, and Alessandro Bressan for helpful discussions. 

We thank Bob Kurucz for developing and maintaining programs and databases without which this work would not be possible. 

DMN acknowledges support from NASA under award Number 80NSSC19K0589, and support from the  Allan C. And Dorothy H.Davis Fellowship. S.H. is supported by the U.S. Department of Energy under Award No. DE-SC0020262, NSF Grant No. AST-1908960, and NSF Grant No. PHY-1914409. RFGW acknowledges support  through the generosity of Eric and Wendy Schmidt by recommendation of the Schmidt Futures program. Y.S.T. is grateful to be supported by the NASA Hubble Fellowship grant HST-HF2-51425.001 awarded by the Space Telescope Science Institute. C.~F.~acknowledges funding provided by the Australian Research Council (Discovery Project DP170100603 and Future Fellowship FT180100495), and the Australia-Germany Joint Research Cooperation Scheme (UA-DAAD). Y.C. acknowledges funding from the ERC Consolida-tor Grant funding scheme (project STARKEY, G.A.n. 615604).

\bibliography{GlobularClustersHelium_PostReferee}

\label{lastpage}
\end{document}